\documentclass[review,12pt,times]{elsarticle}

\usepackage{graphicx}%
\usepackage{multirow}%
\usepackage{amsmath,amssymb,amsfonts}%
\usepackage{amsthm}%
\usepackage{mathrsfs}%
\usepackage[title]{appendix}%
\usepackage{xcolor}%
\usepackage{textcomp}%
\usepackage{manyfoot}%
\usepackage{booktabs}%
\usepackage{algorithm}%
\usepackage{algorithmicx}%
\usepackage{algpseudocode}%
\usepackage{listings}%
\usepackage{soul}%
\usepackage{xr}%
\usepackage{gensymb}%

\journal{Acta Materialia}

\begin{document}

\begin{frontmatter}



\title{Kinetics of Amorphous Defect Phases Measured Through Ultrafast Nanocalorimetry}



\author[UCSB:Mat,UVA:MSE]{W. Streit Cunningham\corref{cor1}}
\author[UA]{Tianjiao Lei}
\author[UCSB:Mat]{Hannah C. Howard}
\author[JHU:Mat,JHU:HEMI]{Timothy J Rupert}
\author[UCSB:Mat]{Daniel S. Gianola\corref{cor1}}

\affiliation[UCSB:Mat]{organization={Materials Department},
            addressline={University of California}, 
            city={Santa Barbara},
            postcode={93106}, 
            state={CA},
            country={USA}}
\affiliation[UVA:MSE]{organization={Department of Materials Science and Engineering },
            addressline={University of Virginia}, 
            city={Charlottesville},
            postcode={22904}, 
            state={VA},
            country={USA}}           
\affiliation[UA]{organization={Department of Metallurgical and Materials Engineering},
            addressline={The University of Alabama}, 
            city={Tuscaloosa},
            postcode={35487}, 
            state={AL},
            country={USA}}
\affiliation[JHU:Mat]{organization={Department of Materials Science and Engineering},
            addressline={Johns Hopkins University}, 
            city={Baltimore},
            postcode={21218}, 
            state={MD},
            country={USA}}
\affiliation[JHU:HEMI]{organization={Hopkins Extreme Materials Institute},
            addressline={Johns Hopkins University}, 
            city={Baltimore},
            postcode={21218}, 
            state={MD},
            country={USA}}

\cortext[cor1]{Corresponding Authors}

\begin{abstract}

Recognition of the role of extended defects on local phase transitions has led to the conceptualization of the \textit{defect phase}, localized thermodynamically stable interfacial states that have since been applied in a myriad of material systems to realize significant enhancements in material properties.
Here, we explore the kinetics of grain boundary confined amorphous defect phases, utilizing the high temperature and scanning rates afforded by ultrafast differential scanning calorimetry to apply targeted annealing/quenching treatments at high rates capable of capturing the kinetic behavior.
Four Al-based nanocrystalline alloys, including two binary systems, Al-Ni and Al-Y, and two ternary systems, Al-Mg-Y and Al-Ni-Y, are selected to probe the materials design space (enthalpy of mixing, enthalpy of segregation, chemical complexity) for amorphous defect phase formation and stability, with correlative transmission electron microscopy applied to link phase evolution and grain stability to nanocalorimetry signatures.
A series of targeted isothermal annealing heat treatments is utilized to construct a Time-Temperature-Transformation curve for the Al-Ni system, from which a critical cooling rate of 2,400 \degree C/s was determined for the grain boundary confined disordered-to-ordered transition.
Finally, a thermal profile consisting of 1,000 repeated annealing sequences was created to explore the recovery of the amorphous defect phase following sequential annealing treatments, with results indicating remarkable microstructural stability after annealing at temperatures above 90\% of the melting temperature.
This work contributes to a deeper understanding of grain boundary localized thermodynamics and kinetics, with potential implications for the design and optimization of advanced materials with enhanced stability and performance.
\end{abstract}

\begin{graphicalabstract}
    \includegraphics[width=1.0\linewidth]{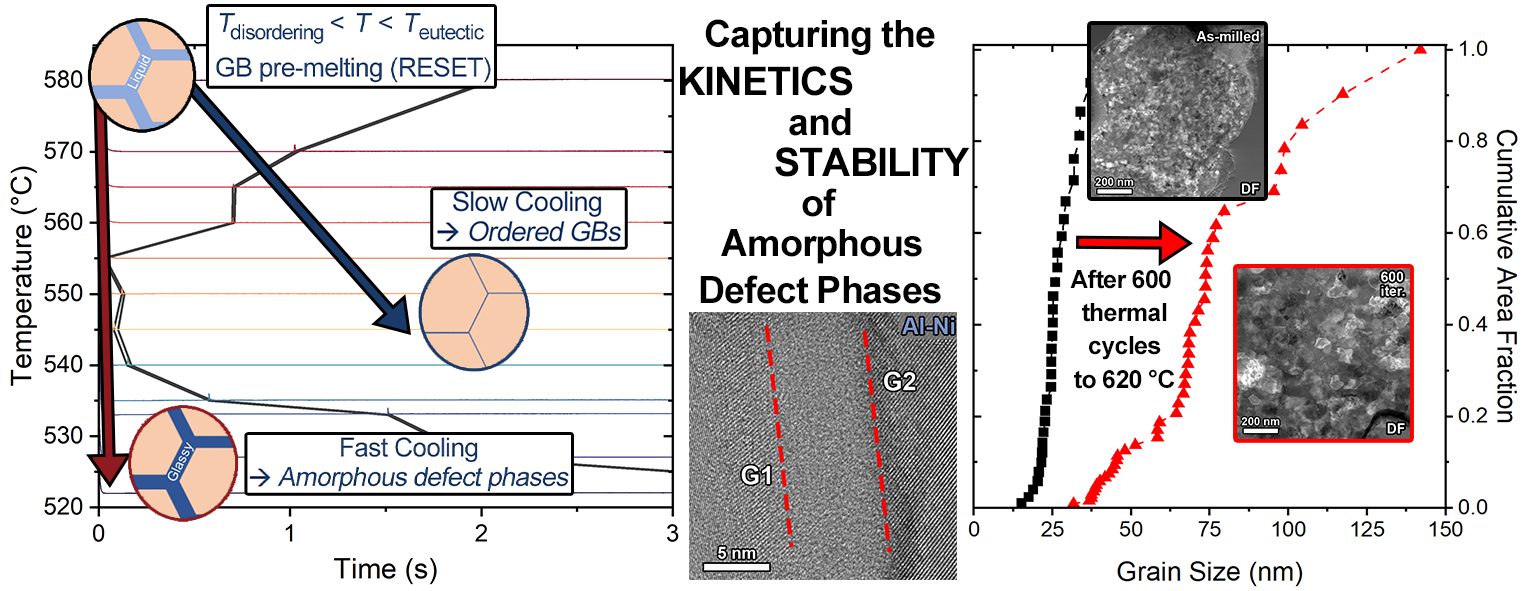}
\end{graphicalabstract}


\end{frontmatter}


\section{Introduction}
\label{sec:introduction}
Nanostructuring, or the purposeful addition of high densities of interfaces, has long been a powerful tool in the metallurgist's toolbox for achieving enhanced alloy properties, driving significant improvements in mechanical behavior (hardness \cite{Suryanarayana1995, KUMAR2003, MEYERS2006}, fracture \cite{Heckman2018, ZHANG2006}), corrosion resistance \cite{LIU2010}, and tolerance under high radiation fluxes \cite{BEYERLEIN2013, CHENG2016}.
Recognition of the role of interfaces, and other extended microstructural defects, has driven the development of so-called \textit{defect phase} diagrams \cite{Korte-Kerzel2022, TEHRANCHI2024, MATSON2024, SHI2011}, in contrast to the bulk phases predicted by typical bulk phase diagrams.
Defect phases, also termed complexions, are structurally and/or chemically distinct configurations localized around extended defects, which can be described by thermodynamic state variables such as the chemical potential, temperature, and composition \cite{DILLON2007, KIRCHHEIM2002}.
While some historical treatments of interfacial defect phases have proposed that these features do not follow the classical treatment of a phase developed by Gibbs \cite{gibbs1878}, a generalized formulation developed by Frolov and Mishin \cite{Frolov2015} allows for the unified treatment of phase equilibria among phases of different dimensionality \cite{CANTWELL2020, CANTWELL2014}.
The flexibility of this formulation has led to the realization of a myriad of unique defect phases along dislocations \cite{HOWARD2025, TURLO2020, KUZMINA2015, SINGH2023}, planar faults \cite{smith2018, titus2016}, and, most relevant to this work, grain boundaries.
\par

Six different grain boundary localized defect phases have been discovered and classified according to thickness, ordering, and composition: sub-monolayer segregation, the intrinsic grain boundary, bi-layer segregation, multi-layer segregation, amorphous grain boundary phase, and wetting films \cite{DILLON2007, Frolov2015PhysRevB, Frolov2013}.
Of particular interest are amorphous grain boundary defect phases, which consist of a disordered phase of finite thickness located at the grain boundaries \cite{MISHIN2010, SCHULER2017, LEI2023, LUO2005, LUO2008, LI2019}.
Nucleation of the amorphous defect phase is a consequence of annealing at elevated temperatures, where a pre-melting ordered-to-disordered (crystalline-to-amorphous) transition can occur at the grain boundaries if the energetic penalty associated with the formation of a confined disordered phase and two new interfaces can be overcome \cite{williams2009, KAPLAN2006, Frolov2013, MELLENTHIN2008, TANG2006}.
Note that this transition differs from the well-established grain boundary wetting phenomena, which occurs within a two-phase (solid + liquid) region of the bulk phase diagram and where the liquid phase completely replaces the grain boundary, separating adjacent grains \cite{straumal2004}.
Instead, grain boundary pre-melting manifests in the single-phase region of the phase diagram at sub-solidus temperatures and involves the formation of a nanoscale liquid-like layer at the grain boundary due to a disjoining potential that stabilizes the film.
Unlike wetting, pre-melting does not require bulk liquid coexistence; instead, it reflects a localized disordering transition driven by interfacial energy minimization.
A phenomenological description of this mechanism, originally developed by Kikuchi and Cahn \cite{kikuchi1980} and illustrated in \textbf{Figure \ref{fig:Figure1}a}, postulates that this energy minimization is achieved when the volumetric free energy barrier for forming an undercooled liquid ($\Delta G$\textsubscript{amorphous}) plus the penalty for forming two crystalline-liquid interfaces is below the grain boundary energy ($\gamma$\textsubscript{GB}):
\begin{equation}
    \label{Eq:GB}
    \gamma_{\text{GB}} > (\gamma_{\text{CL}}^{(1)} + \gamma_{\text{CL}}^{(2)}) + \Delta G_{\text{amorphous}}h
\end{equation}
where $\gamma$\textsubscript{CL} is the crystal-liquid interfacial energy and $h$ is the thickness of the amorphous grain boundary defect phase.
\par

Using doping strategies targeting grain boundary segregation \cite{raabe2014}, amorphous defect phases have been realized in a number of systems, including simple binary alloys such as Cu-Zr and Cu-Hf \cite{SCHULER2017}, Al-Mg, Al-Ni, and Al-Y \cite{LEI2023}, and Ni-W \cite{schuler2018} as well as more complex ternary systems such as Al-Ni-Ce \cite{BALBUS2021, SHIN2022}, Al-Ni-Y \cite{LEI2021, LEI2022}, and Cu-Zr-Hf \cite{grigorian2019, GRIGORIAN2021}.
Drawing a comparison to bulk metallic glasses (BMGs), where formability depends on both thermodynamic and kinetic factors, amorphous defect phase formation in crystalline alloys follows analogous principles: grain boundaries must contain excess solute, and the free energy penalty for creating an undercooled liquid must be minimized.
These selection criteria prioritize dopants with positive enthalpy of segregation and negative enthalpy of mixing \cite{SCHULER2017}.
Similar to BMGs, amorphous defect phase formation is facilitated in more complex alloys, as the increased disorder arising from chemical complexity reduces long range order and promotes the disordered phase \cite{takeuchi2001}.
This behavior has been experimentally observed in Cu-Zr and Cu-Zr-Hf, with the more complex ternary alloy forming thicker amorphous defect phases at significantly slower critical cooling rates \cite{grigorian2019, GRIGORIAN2021}.
\par

Early work on systems containing amorphous defect phases have demonstrated their potential for enhancing thermal stability \cite{schuler2018, BALBUS2021, grigorian2019, grigorian2022}, mechanical properties \cite{khalajhedayati2016, CUNNINGHAM2023, YAN2023}, and radiation tolerance \cite{SCHULER2020, ludy2016}.
The potentially large excess volume of amorphous defect phases has tremendous implications for the elimination of irradiation induced damage, serving as efficient sinks for interstitial/vacancies and defect clusters \cite{ludy2016, aksoy2024}.
Similarly, the ability for amorphous defect phases to annihilate dislocations, along with their ability to reduce grain boundary cracking, has led to significant improvements in mechanical response \cite{khalajhedayati2016}.
As demonstrated in a nanocrystalline ternary alloy (Al-Ni-Ce), the presence of amorphous defect phases at the grain boundaries led to a pronounced increase of strength of 1800 MPa, compared to an average of approximately 700 MPa in a typical Al 7XXX series alloy \cite{BALBUS2021}.
Significantly, this amorphous defect phase containing alloy demonstrated remarkable thermal stability compared to other nanocrystalline alloys, maintaining nanocrystallinity to homologous temperatures (ratio of the temperature to the melting temperature, T\textsubscript{M}) of 0.7T\textsubscript{M}.
This enhanced thermal stability is a consequence of the grain boundary localized disordered state, which, at high temperatures, is the local equilibrium configuration \cite{schuler2018, grigorian2019}.
Application of such features that are thermodynamically stable at high temperatures represents an exciting opportunity to push traditional alloy designs more extreme conditions.
However, while a number of simulation and experimental works have explored the formation of amorphous defect phases \cite{Frolov2015, SCHULER2017}, few works have managed to directly quantify the critical temperatures associated with the phase transition, much less quantify the complex kinetics of the transformation.
\par

In this work, we investigate the kinetic behaviors of amorphous defect phases through targeted nanocalorimetry and microscopy investigations on amorphous defect phase containing alloys.
To examine the influence of increasing chemical complexity on amorphous defect phase formation, we synthesized four different binary and ternary nanocrystalline alloy systems: Al-Ni, Al-Y, Al-Ni-Y, and Al-Mg-Y. 
The high cooling and heating rates from ultrafast differential scanning calorimetry (DSC) are utilized to explore the ordered-to-disordered grain boundary transitions, which are correlated with microstructral information from targeted electron microscopy.
Systematic isothermal annealing runs are performed to establish a Time-Temperature-Transformation (TTT) curve ascribed to the defect phase transitions, which is then employed to determine a critical cooling rate for amorphous defect phase formation.
Finally, a series of repeated annealing sequences is applied to explore the recovery of the grain boundary confined ordered-to-disordered transition.
\par

\section{Materials and methods}
\label{sec:methods}

\subsection{Sample Synthesis}
To fabricate Al-Ni, Al-Y, Al-Ni-Y, and Al-Mg-Y systems, powders of elemental Al (Alfa Aesar, 99.97\%, -100+325 mesh), Ni (Alfa Aesar, 99.9\%, APS 2.2-3.0 $\mu$m) and/or Y (Alfa Aesar, 99.6\%, -40 mesh) or Mg (Alfa Aesar, 99.8\%, -325 mesh) were ball milled for 10 h in a SPEX SamplePrep 8000M high-energy ball mill.
For all systems, the nominal concentration of each solute element is 2 at.\%.
A hardened steel vial and milling media were used with a ball-to-powder weight ratio of 10:1, while 3 wt.\% stearic acid (C18H36O2) was used as a process control agent to prevent excessive cold welding.
The milling process was conducted in a glove-box filled with Ar gas at an oxygen level less than 0.05 ppm to prevent oxidation.
For bulk sample fabrication, the as-milled powders were transferred into a graphite die set and then consolidated into cylindrical bulk pellets with diameter and height of nominally 1 cm using an MTI Corporation OTF-1200X-VHP3 hot press.
The powders were first cold pressed for 10 min under 100 MPa at room temperature to form a green
body, and then hot pressed for 1 h under 100 MPa at 585 \degree C, which is approximately 92\% of the melting
temperature of pure Al.
The heating rate used to reach the target temperature was 10 \degree C/min.
After hot pressing, bulk pellets were slowly cooled down to room temperature with a cooling rate less than 1 \degree C/s by turning off the furnace.
Determination of the average grain sizes for the as-milled powders and consolidated bulk pellets were performed via X-ray diffraction (XRD) scans collected using a Rigaku Ultima III X-ray diffractometer with a Cu K$\alpha$ radiation source operated at 40 kV and 30 mA and using a one-dimensional D/teX Ultra detector.  
Grain size determinations were performed using the Halder-Wagner method with a LaB6 calibration file used as an external standard and  that all systems are nanocrystalline (\textbf{Table \ref{tab:Table1}}, with raw XRD patterns in \textbf{Supplemental Figure 1}).

\begin{table}
    \centering
    \begin{tabular}{c c c c c }
        \textbf{XRD Grain Sizes (nm)} & \textbf{Al-Ni} & \textbf{Al-Y} & \textbf{Al-Ni-Y} & \textbf{Al-Mg-Y}\\
        \hline
        As-milled Powders & $26 \pm 5^*$ & $14 \pm 4^*$ & $20 \pm 2$ & $9 \pm 1$\\
        Consolidated Bulk Pellets & $55 \pm 10$ & $50 \pm 5$ & $54 \pm 6$ & $44 \pm 10$\\
    \end{tabular}
    \caption{Average XRD grain sizes for all four systems: Al-Ni, Al-Y, Al-Ni-Y, and Al-Mg-Y. Values denoted by a $^*$ were estimated from dark-field scanning transmission electron microscopy (STEM-DF) micrographs.}
    \label{tab:Table1}
\end{table}

\subsection{Nanocalorimetry}
\begin{figure} [htbp]
    \centering
        \includegraphics[width=1.0\linewidth]{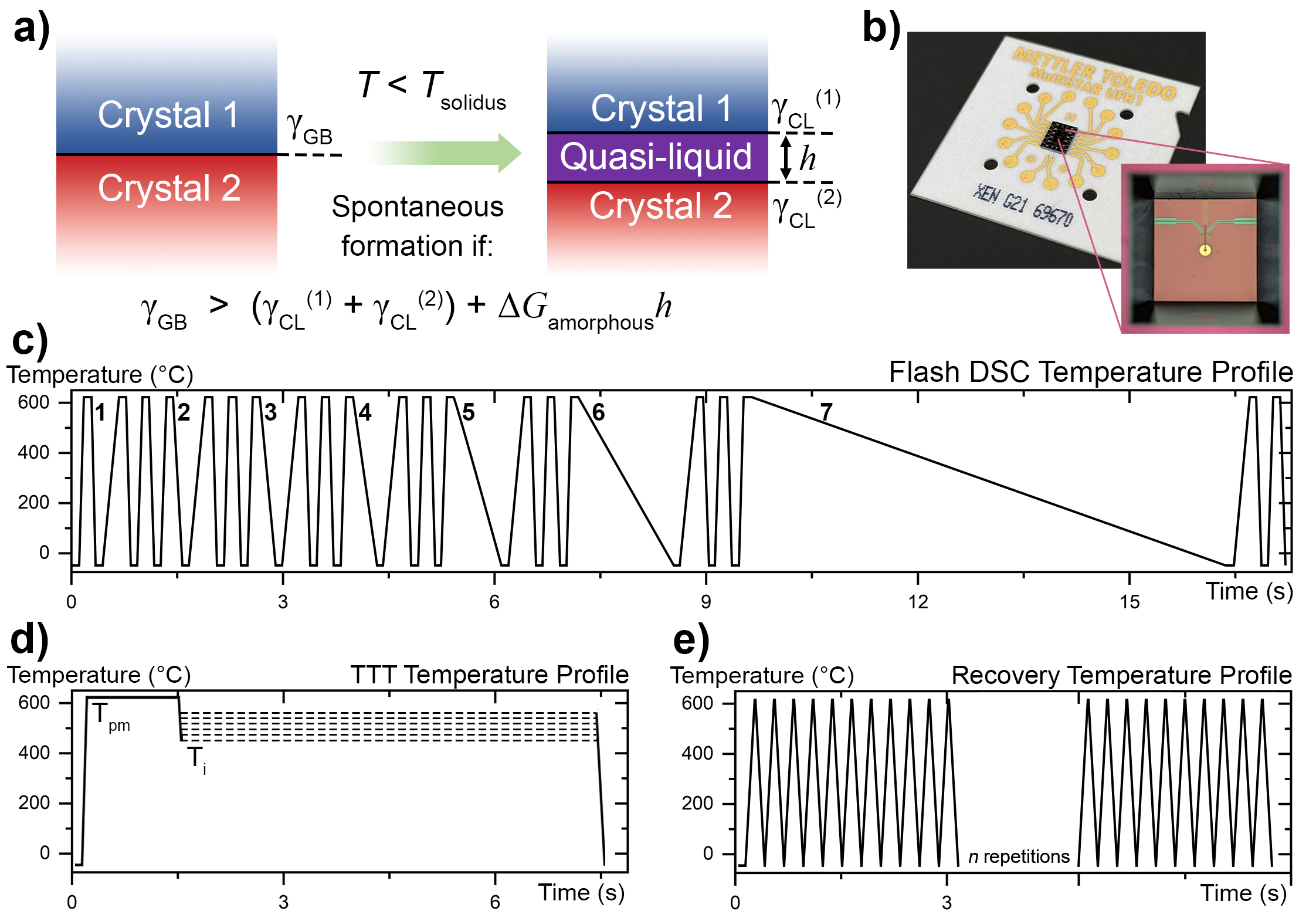}
        \caption{(a) A visual schematic of amorphous defect phase formation. At some temperature below the solidus, T\textsubscript{solidus}, if the interfacial energies of two crystalline-liquid interfaces plus the energetic penalty for an amorphous phase of thickness \textit{h} is below the grain boundary energy, a quasi-liquid, nanoscale amorphous defect phase can form at the grain boundary. (b) The Metter Toledo UFH 1 MEMS chip for the Flash DSC 2+ system, with a magnified image of the chip sensor. (c) Ultrafast DSC thermal profile for exploring amorphous defect phase forming ability. Heating rates are maintained at a constant 3,000 \degree C/s, while seven different rates are used for cooling, ranging from -100 - -10,000 \degree C/s. Two identical heating and cooling rates are used after each cycle for alignment. (d) Ultrafast DSC thermal profile for generating a TTT curve. The sample is rapidly heated to a temperature within the pre-melting regime (T\textsubscript{pm}), held for 1.5 s, and then quenched at -10,000 \degree C/s to a series of isothermal hold temperatures (T\textsubscript{i}) below the pre-melting temperature and held for 6 s. (e) Ultrafast DSC thermal profile for exploring the recovery and stability of amorphous defect phases. The sample is repeatedly heated and quenched (at 5,000 \degree C/s) for n repetitions.}  
        \label{fig:Figure1}
\end{figure}

A high-rate Micro-Electro-Mechanical system (MEMS) based ultrafast DSC from Mettler Toledo, Flash DSC 2+, was used to acquire nanocalorimetry data using as-milled powders, which were manually deposited onto separate UFH 1 MEMS chips for each system.
Because of the difficulties associated with determining the exact mass of micron-scale powders, all DSC data are provided as raw heat flow curves.
A visual of the DSC MEMS chip is provided in \textbf{Figure \ref{fig:Figure1}b}, along with a magnified image of the chip sensor.
Three different thermal profiles were utilized in this study.
The first thermal profile was designed to probe disordered-to-ordered phase transitions at the grain boundaries by systematically varying cooling rates from -100 to -10,000 \degree C/s.
Samples were annealed from -50 \degree C to a temperature below the solidus, with a constant heating rate of 3,000 \degree C/s.
Two identical heating and cooling rates at 10,000 \degree C/s are used after each cycle for alignment.
A graphical representation of this thermal profile is provided in \textbf{Figure \ref{fig:Figure1}c}.
The second thermal profile, in \textbf{Figure \ref{fig:Figure1}d}, was designed to capture the kinetics of the disordered-to-ordered transition associated with amorphous defect phases reverting back to ordered grain boundaries.
Here, samples were annealed to temperatures below the solidus and within the pre-melting regime (T\textsubscript{pm}) for 1.5 s, followed by a rapid quench at -10,000 \degree C/s to an isothermal hold temperature (T\textsubscript{i} for 6 s.
Isothermal hold temperatures were selected from a range of temperatures, 520 - 580 \degree C, below T\textsubscript{pm}.
The third thermal profile was developed to explore the recovery and stability of amorphous defect phases and is shown in \textbf{Figure \ref{fig:Figure1}e}.
Samples were repeatedly heated and cooled (at rates of 5,000 \degree C/s) from -50 \degree C to a temperature below the solidus for \textit{n} iterations, with a maximum number of iterations of 1,000.
\par

\subsection{Electron Microscopy}
As-milled powder samples for transmission electron microscopy (TEM) were prepared by suspending powders in ethanol and pipetting the powder suspension onto carbon lacey grids.
Post-DSC samples were prepared through a standard focused ion beam (FIB) TEM liftout technique using a FEI Helios Dualbeam Nanolab 600 FIB-SEM located in the Microscopy and Microanalysis Facility (MMF) at the University of California, Santa Barbara.
Individual powder particles on the DSC chip were attached to an Omniprobe needle using Pt, moved onto a Cu Omniprobe grid, and subsequently thinned to electron transparency.
TEM micrographs were acquired using a 200 kV Thermo Scientific Talos F200X TEM/STEM also located at the MMF.
The following collection angles were utilized in STEM mode: 9 mrad for bright-field (STEM-BF), 12-20 mrad for dark-field (STEM-DF), and 61-200 mrad for high angle annular dark-field (STEM-HAADF).
\par

\subsection{CALPHAD}
CALPHAD (CALculation of PHAse Diagrams) driven phase diagram calculations for each of the four systems was performed using custom thermodynamic databases.
For the Al-Ni, Al-Y, and Al-Ni-Y systems, a thermodynamic database for the quinary Al-Co-Cr-Ni-Y system from Ref. \cite{Wang2021} was used, while the custom Al-Mg-Y database from Ref. \cite{Chen2024} was used for the Al-Mg-Y system.
For the ternary alloys, a pseudo-binary phase diagram was generated in the Al-rich region across equal composition spaces for the solute species.
\par

\section{Results and Discussion}
\label{sec:results}

\subsection{Selection of the Alloy Systems}

\begin{figure} [htbp]
    \centering
        \includegraphics[width=1.0\linewidth]{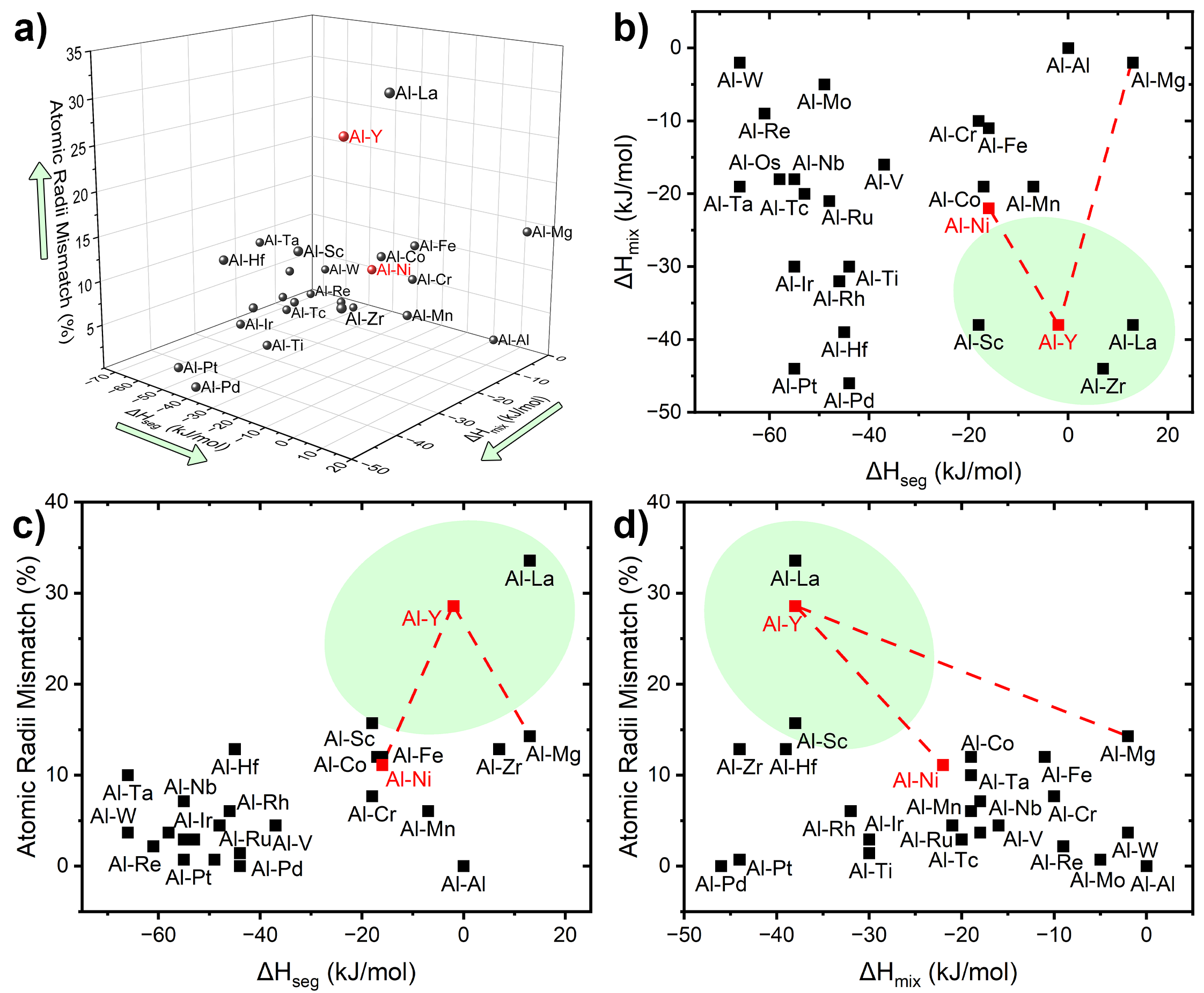}
        \caption{(a) 3D plot of binary Al alloys as a function of atomic radius mismatch, $\Delta H$\textsubscript{seg}, and $\Delta H$\textsubscript{mix} from Refs. \cite{murdoch2013, atwater2012}. Green arrows indicate the directions for increased propensities for amorphous defect phase formation. The binary alloys explored in this study, Al-Ni and Al-Y, are colored in red. (b-d) 2D sections of (a) to facilitate comprehension with regions of optimal parameters for amorphous defect phase formation colored in green. Red dashed lines are drawn between the binary systems to illustrate the two ternary systems explored in this study, Al-Ni-Y and Al-Mg-Y.} 
        \label{fig:Figure2}
\end{figure}

Prior work on amorphous defect phase containing systems by Schuler and Rupert \cite{SCHULER2017} has led to the development of two general materials selection rules for amorphous defect phase formation:
\begin{enumerate}
    \item Sufficient excess solute must be present at the grain boundaries. 
    Propensity for solute segregation at the grain boundaries can be determined through estimation of the enthalpy of segregation, $\Delta H$\textsubscript{seg}, which has been determined for a wide range of binary alloy configurations \cite{murdoch2013}. 
    \item The volumetric free energy penalty for forming an undercooled liquid ($\Delta G$\textsubscript{amorphous}) must be minimized, as it must be energetically favorable for the amorphous structure to form at the grain boundary.
    Given that $\Delta G$\textsubscript{amorphous} can not be directly isolated as a scaling parameter for design, the empirical guidelines for glass forming ability in BMGs \cite{INOUE2000} can be utilized as the primary design parameters, which, as previously discussed, promotes increasing the chemical complexity, utilizing elements with large atomic radii mismatch, and reducing the rate of crystal nucleation by minimizing the enthalpy of mixing, $\Delta H$\textsubscript{mix} (determined for a wide range of binary alloy configurations in Ref. \cite{atwater2012}).
    Additionally, it also important to maintain in mind the factors that promote pre-melting (such as $\gamma_{\text{GB}}$ and $\gamma_{\text{CL}}$ from \textbf{Equation \ref{Eq:GB}}) and understand how these factors vary with composition.    
\end{enumerate}
To represent how changes in atomic radius mismatch, $\Delta H$\textsubscript{seg}, and $\Delta H$\textsubscript{mix} theoretically influence amorphous defect phase formation, we examine a number of binary Al alloys (\textbf{Figure \ref{fig:Figure2}}).
The region illustrating the highest atomic radius mismatch, highest $\Delta H$\textsubscript{seg}, and lowest $\Delta H$\textsubscript{mix} is outlined in green in \textbf{Figure \ref{fig:Figure2}b}, \textbf{c}, and \textbf{d}.
Two binary systems were chosen to cover a range of segregation enthalpies, mixing enthalpies, and atomic size mismatches, with Al-2at.\%Y selected as an ideal amorphous defect phase former and Al-2at.\%Ni selected as a less-than-ideal amorphous defect phase former.
Two ternary alloys were selected to capture the influence of complexity on amorphous defect phase formation, Al-2at.\%Ni-2at.\%Y and Al-2at.\%Mg-2at.\%Y, with one alkaline-earth/transition metal species and one rare earth metal species chosen to maximize the atomic radius mismatch \cite{EGAMI1984, CHENG2011}.
All dopant species have been shown to exhibit limited solubility in Al \cite{okamoto2004, okamoto1998, LIU2006}, implying strong segregation to grain boundaries in Al alloys \cite{devaraj2019, CUNNINGHAM2023, BALBUS2021}.
\par

Importantly, each of the four systems have been previously investigated for amorphous defect phase formation and stability through bulk consolidated pellets \cite{LEI2023, LEI2021}.
Variations in the propensity for amorphous defect phase stability have been observed across both binaries, as evidenced by microstructural characterization with TEM following a slow cooling to room temperature or a fast water quench \cite{LEI2023}.
Amorphous defect phases were present in Al-Y after either cooling process, but only present in Al-Ni after the fast water quench, demonstrating that Al-Y is the most effective amorphous defect phase stabilizer among the investigated binary systems. 
In the two ternary systems, similarly to Al-Y, the presence of amorphous defect phase was determined for both cooling rates, indicating that the addition of chemical complexity promoted amorphous defect phase stability \cite{LEI2021}.
While these prior studies were able to explore the mechanisms underlying amorphous defect phase stability, the analyses were limited to post-annealing characterization of the microstructures, making investigation of the underlying transitions and their critical temperatures difficult.
Here, we aim to quantify kinetics associated with the amorphous defect phase transition within each of the four selected systems.
\par

\subsection{Predicted Bulk Phase Diagrams from CALPHAD}

\begin{figure} [htbp]
    \centering
        \includegraphics[width=1.0\linewidth]{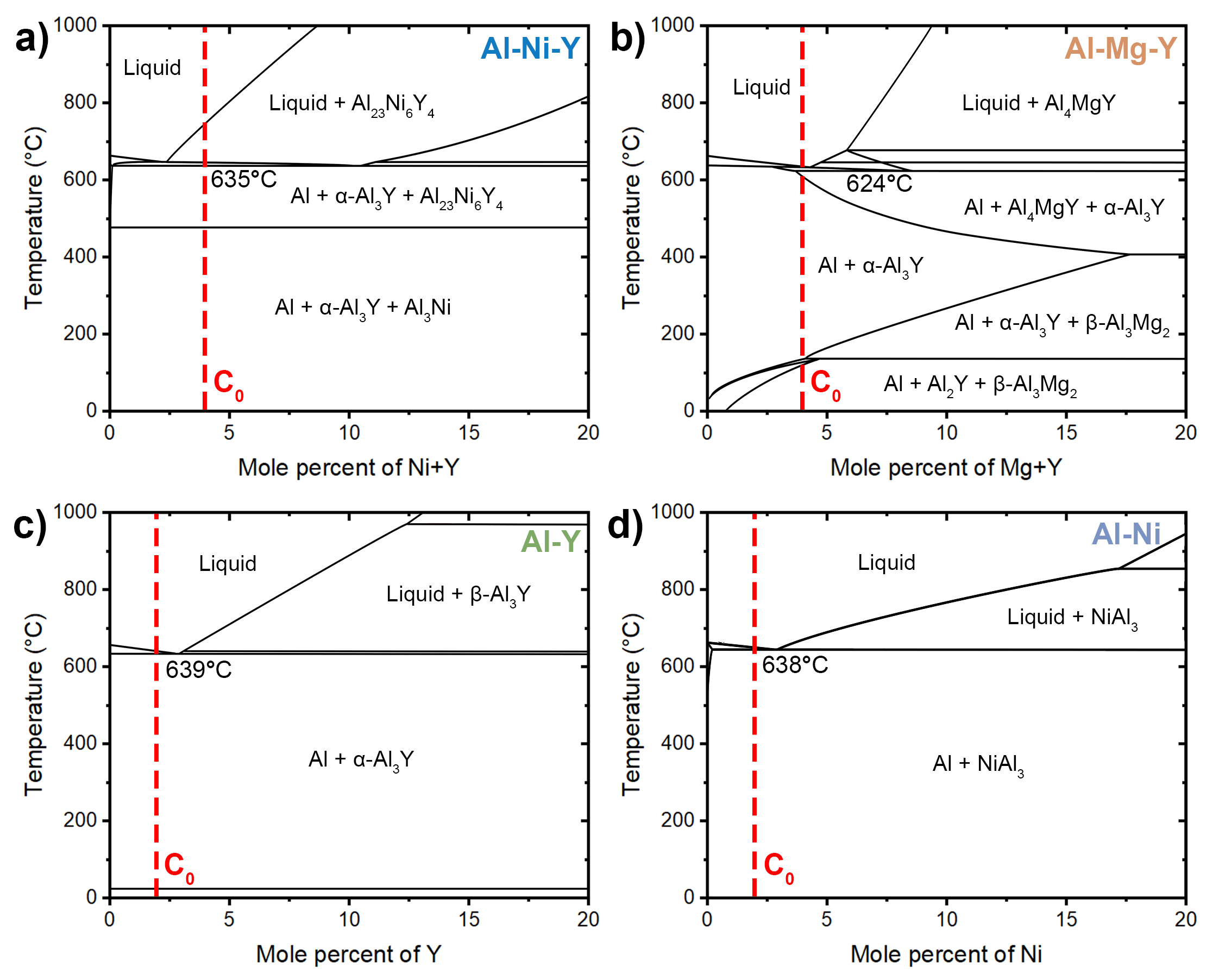}
        \caption{CALPHAD generated pseudo-binary phase diagrams created from vertical cross-sections of the ternary phase diagrams for (a) Al-Ni-Y and (b) Al-Mg-Y, and binary phase diagrams for (c) Al-Y and (d) Al-Ni. Red dashed lines indicate the bulk composition of the alloys, C\textsubscript{0}.} 
        \label{fig:Figure3}
\end{figure}

To determine the optimal solute compositions and temperature range for the nanocalorimetry study, ensuring that the maximum temperature of the thermal profile remained below the solidus, and to predict any potential intermetallic formation during the experiment, bulk phase diagrams (\textbf{Figure \ref{fig:Figure3}}) were calculated for each system using a CALPHAD-based approach.
First, solute concentrations were selected through considerations of the kinetic behavior.
As each system exhibits a eutectic transition, compositions with 2at.\% of solute were chosen to lie close to the eutectic to promote glass formability; a dashed red line is drawn across each phase diagram to indicate the selected solute concentrations.
As the concentration of solute atoms is nominally equivalent in both ternary systems, vertical cross-sections representing the pseudo-binary Al-(Ni,Mg)Y are shown.
Next, to determine the maximum temperature of the thermal profile, we examine the solidus for each system.
However, the solidus varies with respect to composition and thus, eutectic temperatures are instead used as the upper bounds for the thermal profile given that the eutectic temperature is an invariant temperature. 
The pseudo-binary phase diagram of the Al-Ni-Y system is provided in \textbf{Figure \ref{fig:Figure3}a}, and indicates that the eutectic temperature for the Al-Ni-Y system lies at 635 \degree C.
Replacing Ni with Mg promotes a reduction in the eutectic to 624 \degree C (\textbf{Figure \ref{fig:Figure3}b}), while examination of the binary systems, Al-Y and Al-Ni (\textbf{Figures \ref{fig:Figure3}c} and \textbf{d}), indicates eutectic temperatures of 639 \degree C and 638 \degree C, respectively.
\par

Examination of the bulk phase diagrams indicates significant variation in intermetallic formation across all four alloys.
At lower temperatures, the Al-Ni-Y system is expected to consist of predominately FCC Al, with some intermetallic $\alpha-$Al\textsubscript{3}Y and Al\textsubscript{3}Ni, while increasing temperature to nominally 500 \degree C drives the formation of a ternary intermetallic phase, Al\textsubscript{23}Ni\textsubscript{6}Y\textsubscript{4} (monoclinic).
This trend in intermetallic formation for the Al-Ni-Y system, with binary intermetallics forming at lower temperatures followed by ternary intermetallics at higher temperatures, has been well established in the experimental literature \cite{VASILIEV2004, RAGGIO2000, Mika2010, CUNNINGHAM2023}.
Meanwhile, the Al-Mg-Y system shows increasing complexity in intermetallic phases at higher temperatures.
At low temperatures, the Al-Mg-Y systems mimics the Al-Ni-Y system, with a predominately FCC Al phase and some intermetallic binary formation in the form of Al\textsubscript{2}Y and $\beta-$Al\textsubscript{3}Mg\textsubscript{2}.
However, low temperatures of nominally 150 \degree C initiate the formation of the $\alpha-$Al\textsubscript{3}Y intermetallic, while increasing in temperatures to approximately 450 \degree C drives the formation of the ternary Al\textsubscript{4}MgY, first noted in Ref. \cite{zarechnyuk1980}.
Both binary systems are predicted to exhibit fewer variations in intermetallic formation.
In the case of Al-Ni, few intermetallics are expected, with only small quantities of NiAl\textsubscript{3} forming through temperature.
An initial examination of the Al-Y system suggests a similar behavior, with small quantities of the $\alpha-$Al\textsubscript{3}Y intermetallic present; however, annealing at higher temperatures drives the transition between two polymorphic forms of Al\textsubscript{3}Y, from $\alpha-$Al\textsubscript{3}Y to $\beta-$Al\textsubscript{3}Y \cite{LIU2006}.
\par

\subsection{Nanocalorimetry of Four Defect Phase Containing Systems}

\begin{figure} [ht]
    \centering
        \includegraphics[width=1.0\linewidth]{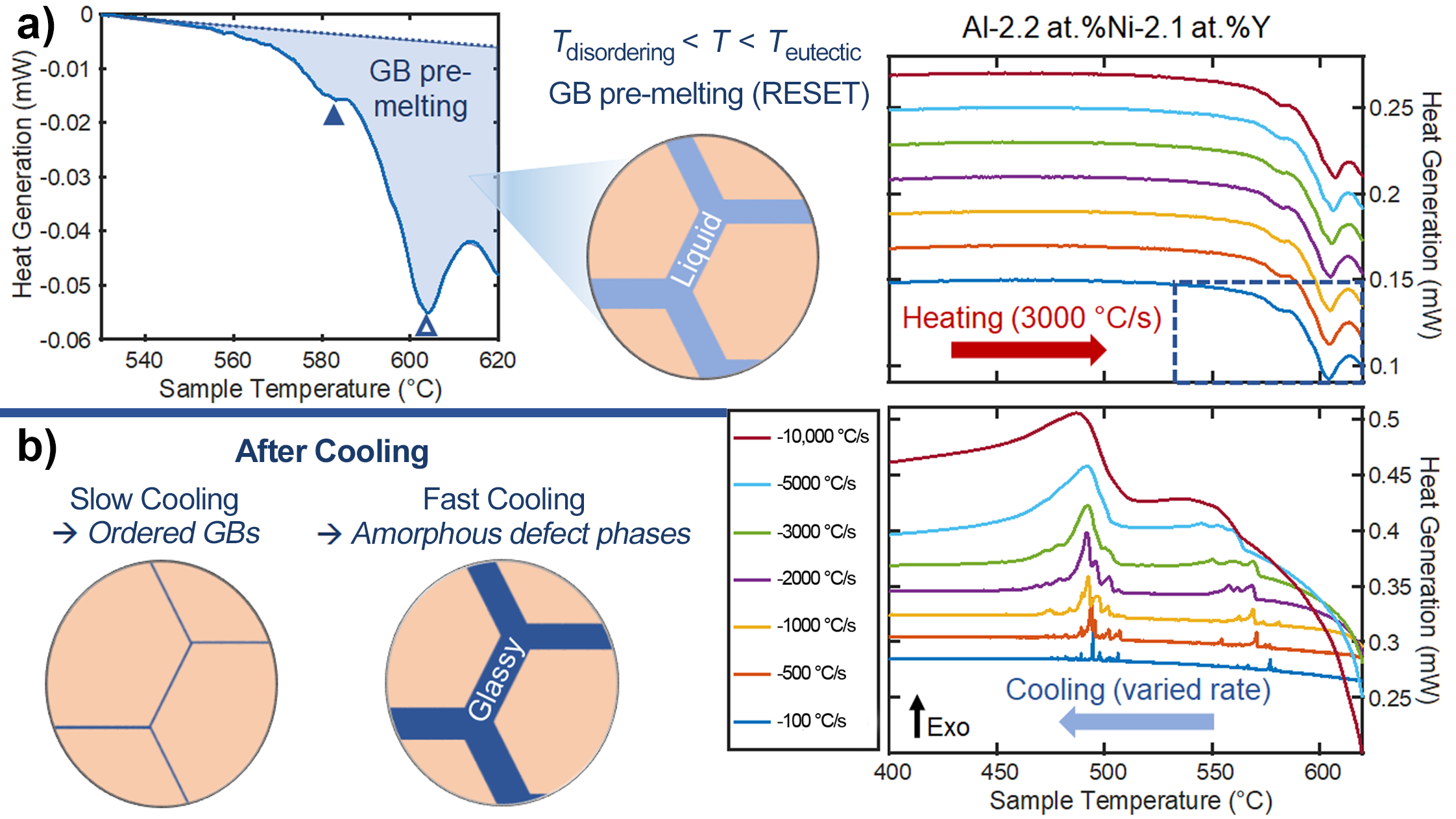}
        \caption{(a) A single heating curve to illustrate the additional endothermic behavior associated with grain boundary pre-melting, which occurs at sub-eutectic temperatures. DSC heating curves for Al-Ni-Y at 3,000 \degree C/s; all DSC curves were acquired after one of the cooling events in (b). (b) DSC cooling curves for Al-Ni-Y at different cooling rates within the range of -100 to -10,000 \degree C/s. Slow cooling rates promote ordering at the grain boundaries, while faster cooling rates kinetically freeze the amorphous defect phase at the grain boundary.} 
        \label{fig:Figure4}
\end{figure}

Here, using Al-Ni-Y as a representative system, we outline a DSC thermal profile that targets the thermodynamics and kinetics associated with a grain boundary confined ordered-to-disordered transition.
The temperature range is bounded by the minimum achievable temperature of the DSC instrument (-50 \degree C) and the eutectic temperature for each alloy system, as any annealing above the eutectic temperature will simply result in bulk melting of the alloy.
Thus, to avoid the eutectic, the DSC thermal profile for Al-Ni-Y was capped at 625 \degree C, 10 \degree C below the eutectic of 635 \degree C determined via CALPHAD (\textbf{Figure \ref{fig:Figure3}a}).
Examination of the DSC heat flows through heating for Al-Ni-Y (\textbf{Figure \ref{fig:Figure4}a}) shows a large endothermic signal near the eutectic that begins at 559 \degree C and continues through the maximum temperature of the thermal profile.
We hypothesize that this signature is related to the grain boundary confined pre-melting transition driven by the minimization of the interfacial energy as a consequence of solute segregation at the grain boundaries.
Its manifestation as an endothermic signature has been previously reported in literature \cite{BALBUS2021}.
Note, however, that a similar endothermic signal would be expected for the dissolution of intermetallic phases in the bulk \cite{li2012, LASA2002, chen2020}, as intermetallics are likely present within the Al-Ni-Y system per the CALPHAD generated phase diagram (\textbf{Figure \ref{fig:Figure3}a}).
However, we propose that the pre-melting transition is the most likely source for this signature, given:
\begin{enumerate}
    \item Three intermetallics are expected to form in Al-Ni-Y at the bulk composition consisting of 2at.\%Ni and 2at.\%Y - $\alpha-$Al\textsubscript{3}Y, Al\textsubscript{3}Ni, and Al\textsubscript{23}Ni\textsubscript{6}Y\textsubscript{4}. 
    Dissolution is not expected for any of these intermetallics within the temperature range of the endothermic signal. 
    The binary phase diagrams for Al-Y and Al-Ni (\textbf{Figure \ref{fig:Figure3}c} and \textbf{d}, respectively) show that these intermetallics are stable up to temperatures of nominally 640 \degree C.
    Furthermore, annealing studies on Al-rich bulk Al-Ni-Y have shown that the Al\textsubscript{23}Ni\textsubscript{6}Y\textsubscript{4} intermetallic is stable within the temperature range of 500 - 610 \degree C \cite{VASILIEV2004}.
    \item The solubility limits for Ni and Y in Al are extremely low, around 0.1at.\% at near eutectic temperatures \cite{SHAO2023, LIU2006}, promoting intermetallic formation over a solid solution.
\end{enumerate}
Furthermore, we note that repetition of the annealing sequence (as demonstrated in the DSC heat flow curves in the right panel of \textbf{Figure \ref{fig:Figure4}a}) did not yield any appreciable change in subsequent curves, suggesting that no significant microstructural changes are occurring in the sample following the annealing sequence.
This behavior, in addition to the lack of exothermic signatures at lower temperatures (below 500 \degree C) that are characteristic of solute diffusion towards the grain boundaries \cite{BALBUS2021}, also indicates that solute segregation at the grain boundaries occurred prior to DSC analysis and during ball milling.
\par

As amorphous defect phases are at a local equilibrium within the pre-melting regime, cooling must be performed at sufficiently high rates to kinetically freeze in the grain boundary defect phase state at lower temperatures.
To capture this behavior in a single sample, cooling rates were varied from -100 to -10,000 \degree C/s; the DSC curves for each cooling rate in Al-Ni-Y are provided in \textbf{Figure \ref{fig:Figure4}b}.
When cooling rates are low, sufficient time is allotted to enable ordering at the grain boundaries, as evidenced by the sharper exothermal peaks at -100 \degree C/s.
At higher cooling rates, the rapid thermal quench suppresses atomic rearrangement, preventing structural ordering and locking the grain boundary into a frozen disordered (glassy) configuration. 
This kinetic freezing behavior and the associated glass transition appear in DSC curves as a broad exothermal peak (reflecting delayed energy release during constrained atomic mobility) and a distinct step-like change in heat flow, respectively.
Such signatures are characteristic of a glass transition \cite{janovszky2022} and are clearly present in the DSC cooling curve at -10,000 \degree C/s.
\par

\begin{figure} [ht]
    \centering
        \includegraphics[width=1.0\linewidth]{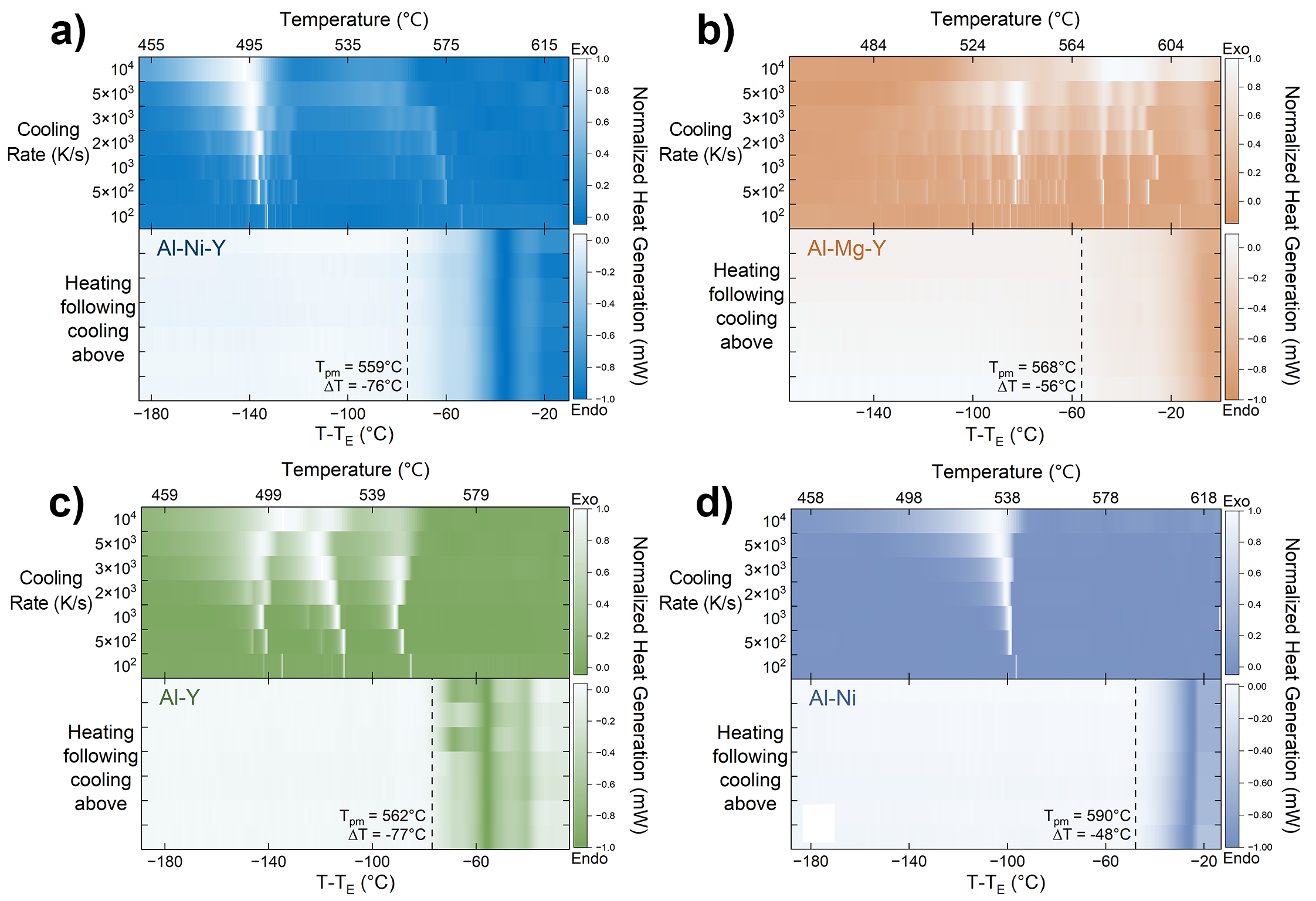}
        \caption{Normalized DSC heat flow curves for seven different cooling rates (-100 - -10,000 \degree C/s) and the heating curves following rapid cooling (maintained at 3,000 \degree C/s) for all four systems: (a) Al-Ni-Y, (b) Al-Mg-Y, (c) Al-Y, and (d) Al-Ni. Temperatures are normalized by the eutectic predicted through CALPHAD and heat flow curves are normalized to max values.} 
        \label{fig:Figure5}
\end{figure}

DSC heat flow curves using the thermal profile outlined above (shown in \textbf{Figure \ref{fig:Figure1}c}) were collected for each of the four Al-based alloys, with results provided in \textbf{Figure \ref{fig:Figure5}}.
Analysis of the heating curves shows little change in the shape of the curves across each system regardless of the previous cooling rate, confirming that the pre-melting regime in each system does not strongly depend on prior thermal history.
Determination of the onset of the pre-melting regime (defined by the pre-melting temperature, T\textsubscript{pm}) is performed by identifying the inflection point of the heating curve; the onset temperature for each system, in order of lowest to highest temperature, is 559 \degree C in Al-Ni-Y, 562 \degree C in Al-Y, 568 \degree C in Al-Mg-Y, and 590 \degree C in Al-Ni.
Ordering each system by $\mid$$\Delta$T$\mid$, or difference in the onset of the pre-melting regime relative to the eutectic ($\mid$$\Delta$T$\mid$ = $\mid$T - T\textsubscript{E}$\mid$) gives: 77 \degree C in Al-Y, 75 \degree C in Al-Ni-Y, 56 \degree C in Al-Mg-Y, and 48 \degree C in Al-Ni (values are summarized in \textbf{Table \ref{tab:Table2}}).
Of all systems, Al-Ni features the smallest pre-melting regime, beginning at 590 \degree C with a $\mid$$\Delta$T$\mid$ of only 48 \degree C.
Every other system features a $\mid$$\Delta$T$\mid$ of at least 50 \degree C, with Al-Y exhibiting the highest $\mid$$\Delta$T$\mid$ of 77 \degree C.
\par

Examination of our results, while maintaining the generalized rules for amorphous defect phase formation in mind, clarifies the source of the trends in $\mid$$\Delta$T$\mid$.
Considering the two binary systems in this study, Al-Ni and Al-Y, Al-Ni has the lowest $\Delta H$\textsubscript{seg} (-16 kJ/mol, compared to -2 kJ/mol in Al-Y), the highest $\Delta H$\textsubscript{mix} (-22 kJ/mol, compared to -38 kJ/mol in Al-Y), and the lowest atomic radius mismatch (10\%, compared to 29\% in Al-Y \cite{laws2015}).
Furthermore, it is well established that selection of solutes with larger atomic radii drives a reduction in $\gamma_{\text{GB}}$ \cite{HUANG2019} and in this metric, Al-Y is also the most favorable system.
Indeed, in every metric for amorphous defect phase forming ability proposed by Schuler and Rupert \cite{SCHULER2017}, Al-Ni scores below Al-Y, and this trend is reflected in $\mid$$\Delta$T$\mid$, with Al-Ni exhibiting a $\mid$$\Delta$T$\mid$ of 48 \degree C relative to Al-Y's $\mid$$\Delta$T$\mid$ of 77 \degree C.
We note that $\mid$$\Delta$T$\mid$ behaves analogously to the supercooled liquid region in metallic glasses (defined as the difference between the crystallization temperature and the glass transition temperature, T\textsubscript{X} - T\textsubscript{g}), with larger supercooled liquid regions indicating higher thermal stability and a greater potential for forming metallic glasses \cite{miller2007}.
This behavior suggests that $\mid$$\Delta$T$\mid$ is a useful experimental metric for amorphous defect phase forming ability: the lower the $\mid$$\Delta$T$\mid$, the higher the necessary temperature, or driving force, for amorphous defect phase formation.
\par

\begin{table}
    \centering
    \begin{tabular}{c c c }
        \textbf{Alloy Systems} & \textbf{T\textsubscript{pm} (\degree C)} & \textbf{$\mid$$\Delta$T$\mid$ (\degree C)}\\
        \hline
        Al-Ni & 590 & 48\\
        Al-Y & 562 & 77\\
        Al-Ni-Y & 559 & 76\\
        Al-Mg-Y & 568 & 56\\
    \end{tabular}
    \caption{Summary of the onset temperatures for the pre-melting regime and $\mid$$\Delta$T$\mid$ = $\mid$T - T\textsubscript{E}$\mid$ for all alloy systems.}
    \label{tab:Table2}
\end{table}

While direct comparisons between the two binary systems is straightforward, a direct one-to-one comparison of the two ternary systems, Al-Ni-Y and Al-Mg-Y, is not as simple. 
Performing an extrapolation between the base binary systems (Al-Ni and Al-Mg) of the ternaries suggests that Al-Mg-Y would be the better amorphous defect phase former given Al-Mg's higher $\Delta H$\textsubscript{seg} at 13 kJ/mol and atomic radii mismatch at 14\%.
However, Al-Mg has a much higher $\Delta H$\textsubscript{mix} relative to Al-Ni (-2 kJ/mol vs. -22 kJ/mol, respectively), and experimental evidence from Lei et al. \cite{LEI2023} in bulk Al-Mg alloys showed no evidence of amorphous defect phase formation under any thermal processing conditions, thus supporting the inverse conclusion that Al-Ni-Y is the relative best amorphous defect phase forming system.
Comparison of the $\mid$$\Delta$T$\mid$ between each alloy supports this conclusion, with Al-Mg-Y exhibiting a lower $\mid$$\Delta$T$\mid$ at 56 \degree C compared to Al-Ni-Y at 75 \degree C.
Nonetheless, estimates of $\Delta H$\textsubscript{mix} and particularly $\Delta H$\textsubscript{seg} for a ternary system are not a simple superposition of the binaries, as extrapolation from the binary systems does not correctly account for ternary interactions.
Further comparison between these ternary systems can be performed through examination of experimental evidence from literature; sintering curves for the bulk ternary systems indicate an earlier onset temperature for activated sintering in Al-Mg-Y, suggesting an earlier onset of amorphous defect phase formation \cite{LEI2021}.
Finally, note that the addition of chemical complexity in the form of an additional solute species drives clear improvements in amorphous defect phase forming ability, as evidenced by the increase in $\mid$$\Delta$T$\mid$ in Al-Ni-Y (75 \degree C) relative to Al-Ni (48 \degree C).
Here, as in BMGs, the additional chemical complexity reduces long range order, reduces $\gamma_{\text{GB}}$, and promotes amorphous defect phase formation \cite{ZHANG2024, takeuchi2001}.
\par

Examination of the DSC cooling curves performed at different cooling rates in the range of -100 to -10,000 \degree C/s (\textbf{Figure \ref{fig:Figure5}}) reveals a similar behavior in each of the four systems.
At low cooling rates, reordering of the grain boundary is possible, resulting in sharp exothermal peaks.
Increasing in cooling rate results in broadening of the exothermal signals, creating an amorphous response as ordering of the boundaries is suppressed.
Interestingly, each of the four systems exhibits a different number and distribution of exothermal signals.
Three distinct possibilities exist for the difference in number of exothermal peaks between each system:
\begin{enumerate}
    \item \textbf{Grain boundary character} is intimately connected to grain boundary energy \cite{OLMSTED2009}; as amorphous defect phase formation is reliant on overcoming the intrinsic grain boundary energy, grain boundaries with specific grain boundary character could be more prone to amorphous defect phase formation \cite{GARG2021}.
    However, it is unlikely that the variations in exothermal signals in this case are related entirely to grain boundary character.
    Distinct, sharp exothermal signals would likely require a distinct population of specific grain boundary characters arising from a heavily textured microstructure, yet all microstructures for this study contain equiaxed nanocrystalline grains (as-milled microstructure of nanocrystalline Al-Y powders are provided in \textbf{Figure \ref{fig:Figure6}} which are expected to contain a broad distribution of grain boundary characters \cite{bober2016}).
    \item It is well established that the presence of extremely \textbf{small particle sizes} can drive changes in the melting and solidification of materials due to the increased surface-to-volume ratio relative to bulk materials \cite{VANFLEET1995, skripov1981}.
    In the case where a distribution of small particles sizes is present, it would not be unexpected to observe appreciable variations in exothermal peaks in the DSC cooling curves.
    However, this behavior is typically only observed for extremely small particle sizes (below 50 nm), and as such, is not expected to be the main driving force for the variations in exothermal peaks observed in this work as average particle sizes for each of the four systems are in the range of 2 to 50 $\mu$m and thus well above 50 nm.
    \item Continuous heating and cooling runs can drive microstructural changes, enabling the formation of \textbf{additional intermetallic features}. 
    Examination of the CALPHAD-predicted phase diagrams in \textbf{Figure \ref{fig:Figure3}} shows that three out of the four systems, Al-Y, Al-Ni-Y, and Al-Mg-Y, are predicted to form various intermetallics with increasing temperature.
    Only one system, Al-Ni (\textbf{Figure \ref{fig:Figure3}d}), is expected to form only a single intermetallic.
    Thus, the single exothermal peak in the DSC cooling curves in Al-Ni (\textbf{Figure \ref{fig:Figure5}d}) can be attributed to the grain boundary confined disordered-to-ordered transition.
    For the other systems, increasing intermetallic complexity in the CALPHAD-predicted phase diagrams generally results in increased complexity in the DSC cooling curves: in Al-Ni-Y an additional phase transition at nominally 500 \degree C associated with the ternary Al\textsubscript{23}Ni\textsubscript{6}Y\textsubscript{4} intermetallic is present, in Al-Mg-Y various phase transitions associated with Al\textsubscript{4}MgY and $\beta-$Al\textsubscript{3}Mg\textsubscript{2} are present at temperatures above 500 \degree C, and in Al-Y, the transition between $\alpha-$Al\textsubscript{3}Y and $\beta-$Al\textsubscript{3}Y lies close to the eutectic.
\end{enumerate}
\par

\subsection{Microstructural Variations Through Temperature}

\begin{figure} [ht]
    \centering
        \includegraphics[width=0.75\linewidth]{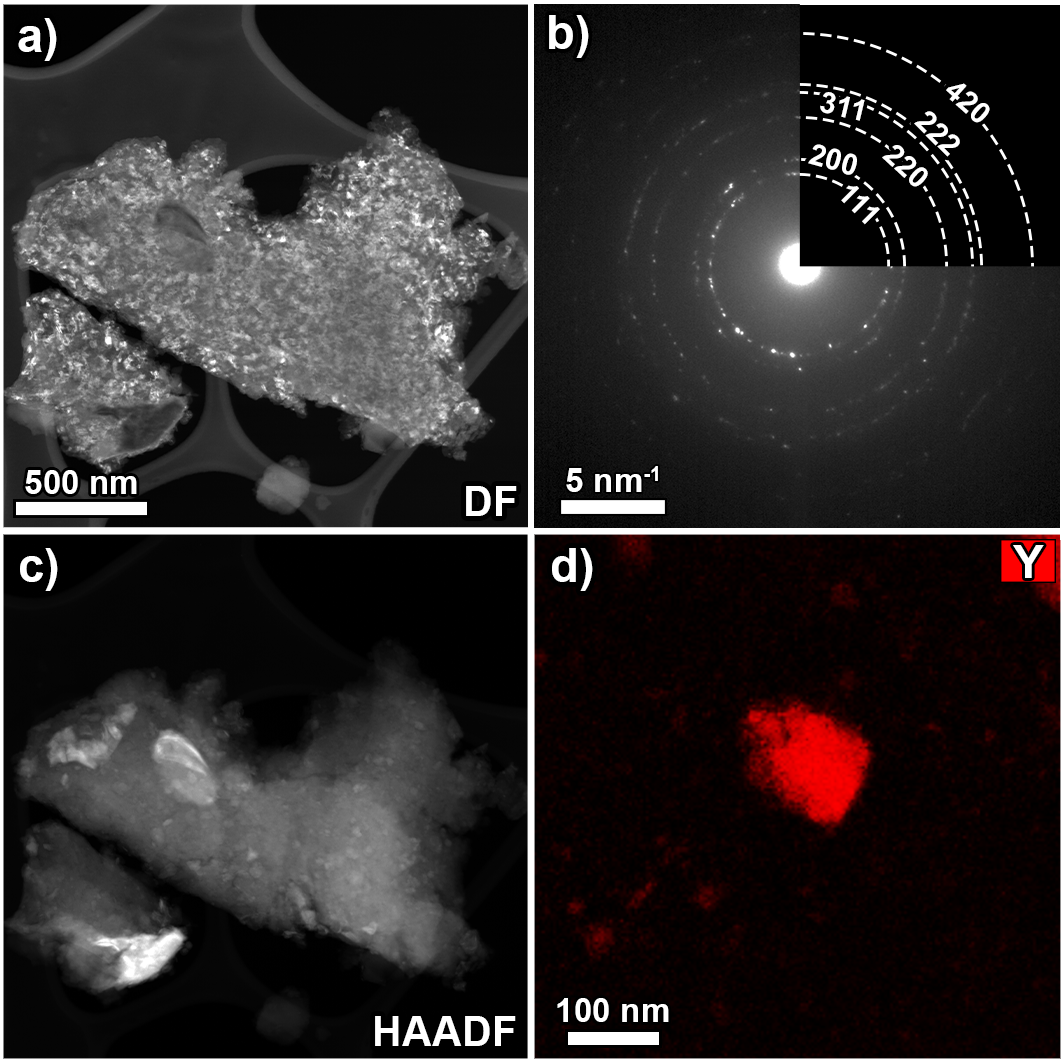}
        \caption{(a) STEM-DF micrograph of as-milled powders of Al-Y. (b) Selected area diffraction pattern of the nanocrystalline Al-Y in (a), with matching diffraction rings corresponding to FCC Al. (c) STEM-HAADF micrograph of as-milled powders of Al-Y. (d) A higher magnification complementary STEM-EDS map of Y (red). Micrographs in (a) and (c) utilize the same scale-bar.} 
        \label{fig:Figure6}
\end{figure}

The variability in exothermal DSC peaks across each of the four systems reflects the differences in intermetallic formation upon annealing of the microstructure.
To link the presence of intermetallics with DSC signatures and examine their influence on the microstructure, TEM samples were prepared from a representative Al-Y sample before and after a single ultrafast DSC run within the pre-melting regime followed by a cooling rate of -3000 \degree C/s.
Beginning with the initial state, as-milled microstructures of Al-Y are provided in \textbf{Figure \ref{fig:Figure6}}, with the STEM-DF micrograph (\textbf{Figure \ref{fig:Figure6}a}) exhibiting a clear initial nanocrystalline structure. 
This nanocrystallinity is also evident in the selected area diffraction (SAED) pattern in \textbf{Figure \ref{fig:Figure6}b}, which shows the distinct ring pattern characteristic of nanocrystalline metals \cite{williams1996}.
Indexing of this SAED pattern matches the expected peaks for face-centered cubic (FCC) Al.
A corresponding STEM-HAADF micrograph (\textbf{Figure \ref{fig:Figure6}c}), where the contrast is directly related to atomic number \cite{williams1996}, indicates the presence of intermetallics, a likely consequence of elevated temperatures during milling, with STEM-EDS maps of Y (\textbf{Figure \ref{fig:Figure6}d}) corroborating the presence of intermetallics.
Notably, despite the presence of intermetallics, the microstructure remained nanocrystalline and thus, sufficient solute remained to segregate towards the grain boundaries and thermodynamically stabilize the microstructure.
While we do not directly demonstrate the presence of grain boundary segregation here, this mechanism has been well established for these alloy systems (Al-Ni-Y \cite{CUNNINGHAM2023, LEI2021, LEI2022}, Al-Mg-Y \cite{LEI2021}, Al-Ni and Al-Y \cite{LEI2023}).
\par

\begin{figure}
    \centering
        \includegraphics[width=1.0\linewidth]{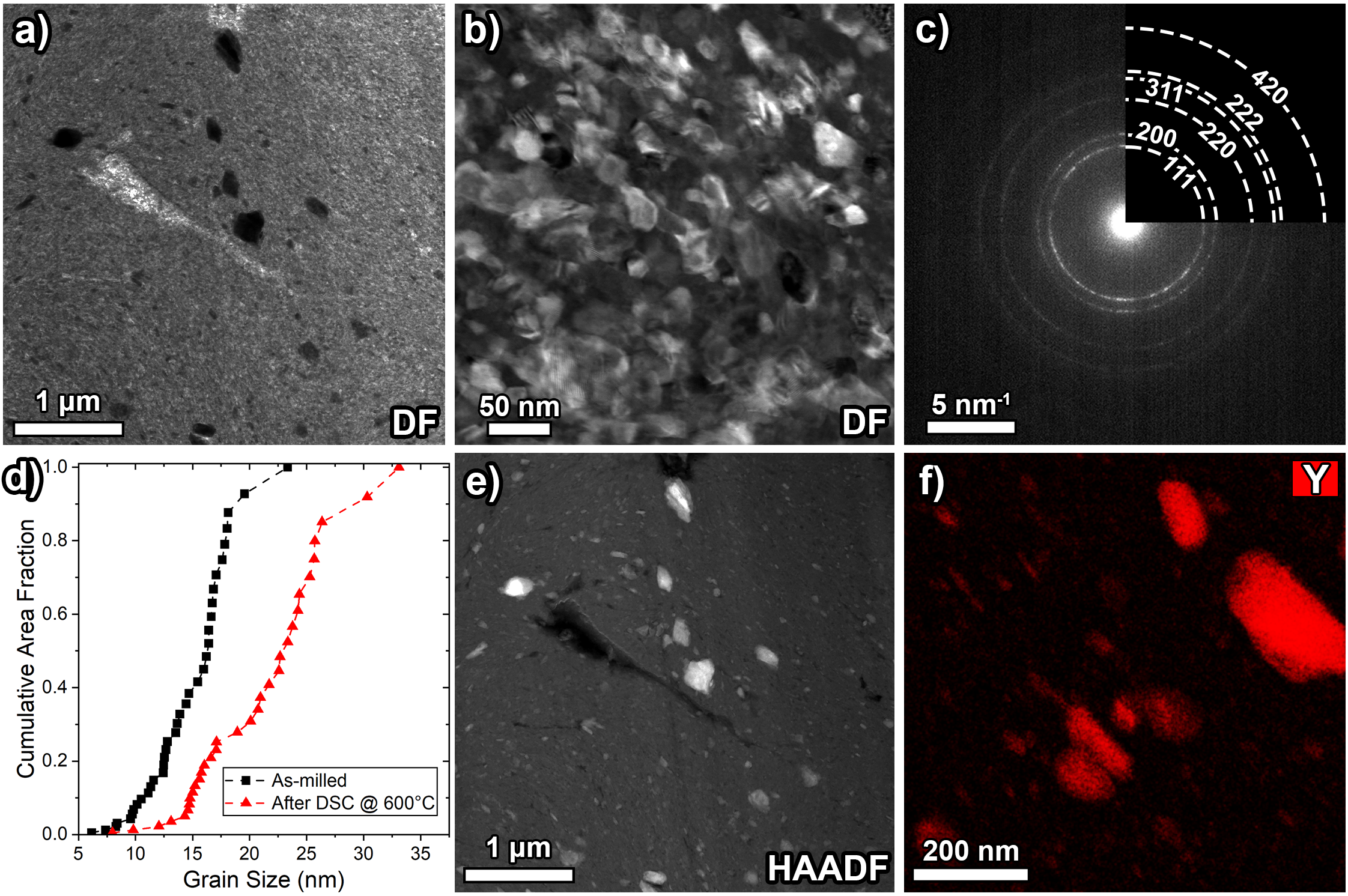}
        \caption{(a) STEM-DF micrograph of Al-Y following a single ultrafast DSC annealing run at 3,000 \degree C/s to 600 \degree C followed by a -3,000 \degree C/s quench. (b) A higher magnification STEM-DF micrograph showing a clear nanocrystalline structure. (c) Selected area diffraction pattern of the nanocrystalline Al-Y in (a), with matching diffraction rings corresponding to FCC Al. (d) Cumulative grain size distributions for the as-milled Al-Y from \textbf{Figure \ref{fig:Figure6}} (black) and post-annealed Al-Y (red). (e) Corresponding STEM-HAADF micrograph for the region in (a). (f) A higher magnification complementary STEM-EDS map of Y (red).} 
        \label{fig:Figure7}
\end{figure}

Next, a FIB lift-out of a single powder was performed following a single DSC anneal to 600 \degree C and subsequent thinning to electron-transparency.
A qualitative examination through STEM-DF of the microstructure post-annealing (\textbf{Figure \ref{fig:Figure7}a} and \textbf{b}) suggests the continued presence of the nanocrystalline structure, demonstrating the inherent thermal stability of such amorphous defect phase containing systems.
Further evidence of nanocrystallinity is present in the SAED pattern (\textbf{Figure \ref{fig:Figure7}c}), which exhibits the same ring pattern present in \textbf{Figure \ref{fig:Figure6}b} with little to no variation.
To quantitatively determine any microstructural changes, grain size distributions were measured from STEM-DF micrographs of both conditions (\textbf{Figure \ref{fig:Figure7}d}).
Average grain sizes are shown to increase subtly from $13.8 \pm 3.8$ nm in the as-milled Al-Y to $19.4 \pm 5.8$ nm after annealing, a strikingly small increase despite the rapid annealing at a very high homologous temperature of 0.94T\textsubscript{M}.
Furthermore, intermetallics are still clearly present, as observed through STEM-HAADF and STEM-EDS (\textbf{Figure \ref{fig:Figure7}e} and \textbf{f}), but not at sufficient fractions to destabilize the microstructure.
Thus, despite the presence of intermetallics and the inherent instability of nanocrystalline grain sizes, our nanocrystalline Al alloys are exceptionally stable, maintaining nanocrystallinity at 0.94T\textsubscript{M} due to enhanced grain boundary stability from the presence of amorphous defect phases.
\par

\begin{figure} [ht]
    \centering
        \includegraphics[width=0.5\linewidth]{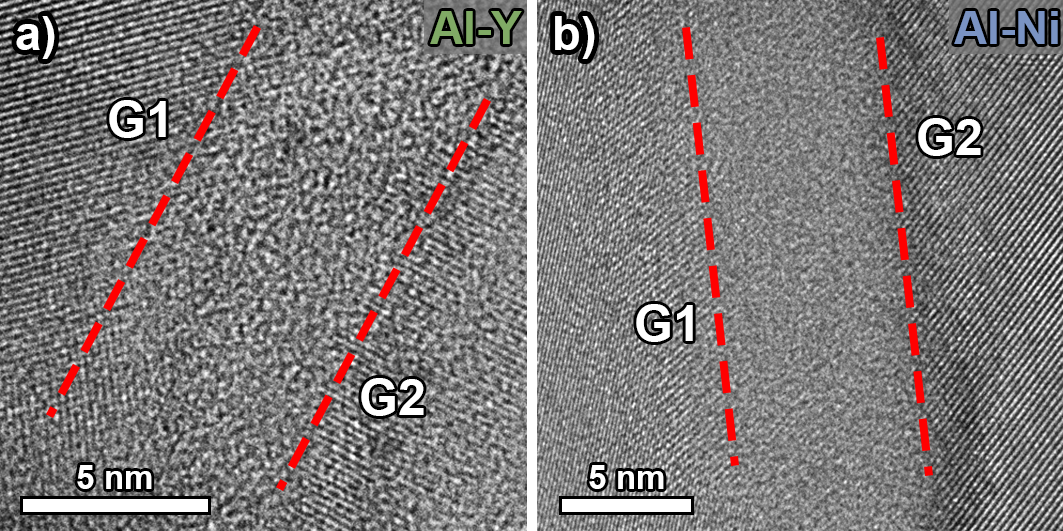}
        \caption{High resolution TEM micrographs of amorphous defect phases in bulk consolidated samples in (a) Al-Y after a slow quench and (b) Al-Ni after quenching in water. Red lines outline the amorphous defect phases and adjacent grains are denoted by "G1" and "G2."}
        \label{fig:Figure8}
\end{figure}

Thermal signatures of amorphous defect phase formation should be accompanied by microstructural evidence, provided the cooling rate is sufficiently fast.
Here, we provide direct visual evidence of amorphous defect phase formation through high resolution TEM of bulk consolidated powder samples, either naturally cooled to room temperature or rapidly quenched in water, as described in Ref. \cite{LEI2023}.
We elected to perform this analysis on bulk consolidated samples that exhibit slightly larger grain sizes (50-100 nm) relative to powders (15-30 nm), which facilitates TEM characterization of grain boundary features.
As TEM samples are nominally 100 nm in thickness, imaging of the bulk consolidated samples facilitates imaging along a single grain, as opposed to imaging through overlapping grains in the powders.
Note that prior literature has demonstrated the presence of amorphous defect phases in all four systems explored in this paper (Al-Ni-Y in Ref. \cite{LEI2021}, Al-Mg-Y in Ref. \cite{LEI2022}, and Al-Ni and Al-Y in Ref. \cite{LEI2023}), so here we limit our examination to the two binary systems, Al-Y and Al-Ni.
As shown in \textbf{Figure \ref{fig:Figure8}}, both binary systems form an amorphous defect phase upon cooling, with some variation in the cooling rate required to maintain the amorphous defect phase.
For Al-Y (\textbf{Figure \ref{fig:Figure8}a}), amorphous defect phases were observed after naturally cooling the sample to room temperature, while amorphous defect phases were only observed in Al-Ni (\textbf{Figure \ref{fig:Figure8}b}) after water quenching, which has a cooling rate of nominally -1,000 \degree C/s \cite{tiryakioglu1998}.
This behavior is a clear reflection of the stronger propensity for amorphous defect phase formation in Al-Y, with a $\mid$$\Delta$T$\mid$ of 77 \degree C relative to Al-Ni, where $\mid$$\Delta$T$\mid$ is 48 \degree C.
Finally, the amorphous defect phases, outlined by the red dashed line, have thicknesses as high as 5-10 nm and are comparable in length-scale to previously published literature on these binary systems \cite{LEI2023}.
We note that thinner amorphous defect phases are observed as well.
\par

\subsection{Measuring a Defect Phase Time-Temperature-Transformation Curve}

\begin{figure} [htbp]
    \centering
        \includegraphics[width=0.5\linewidth]{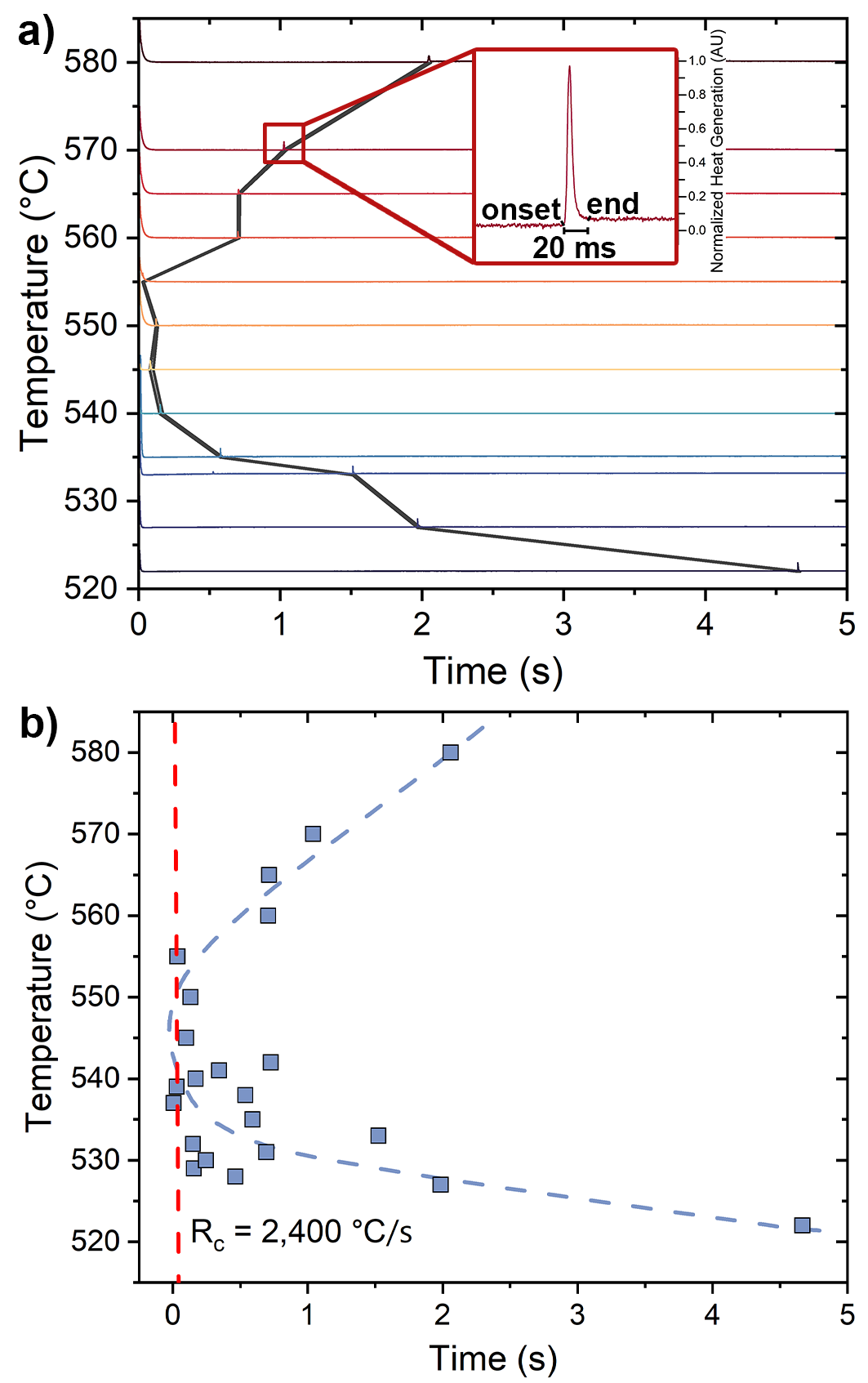}
        \caption{(a) Isothermal traces at various temperatures between 520 - 580 \degree C for Al-Ni. A trace of the TTT curve is formed by outlining the DSC signal; an expanded view of the DSC signal is provided as an inset. (b) A complete scatter plot of the TTT curve for Al-Ni consisting of every collected data point. A dashed red line indicates the cooling curve generated to estimate the critical cooling rate, or the minimum cooling rate necessary to maintain the disordered amorphous defect phase state. For Al-Ni, this critical cooling rate is -2,400 \degree C/s.}
        \label{fig:Figure9}
\end{figure}

We next measure the kinetics of the disordered-to-ordered transition associated with amorphous defect phases by generating a Time-Temperature-Transformation (TTT) curve specifically for the Al-Ni system, which exhibited only a single exothermal DSC signal upon cooling.
A thermal profile with multiple isothermal segments, \textbf{Figure \ref{fig:Figure1}d}, was developed to construct the TTT curve; this thermal profile was devised following methodologies from prior literature utilizing ultrafast DSC to construct TTT curves \cite{DUAN2024, YANG2021, Pogatscher2014, neuber2020}.
The sample was annealed within the pre-melting regime (625 \degree C for Al-Ni) for 1.5 s, and then rapidly quenched at -10,000 \degree C/s to an isothermal hold temperature (T\textsubscript{i}) within the range of 520 - 580 \degree C.
As expected for the Al-Ni system, only a single rapid (20 ms) exothermal DSC signal occurred throughout each isothermal hold (shown in the inset of \textbf{Figure \ref{fig:Figure9}a}).
By measuring the onset and end of each exothermal DSC signal, a TTT curve associated with the grain boundary confined transition was constructed (\textbf{Figure \ref{fig:Figure9}a}).
Creation of a TTT curve for this disordered-to-ordered transition represents, to the authors' knowledge, the first direct construction from calorimetry data of a TTT curve for a grain boundary confined phase transition, as prior literature on the matter have typically inferred approximate TTT curves through post-mortem electron microscopy \cite{GRIGORIAN2021, CANTWELL2016}.
As expected for a TTT curve, the nose of the curve, where the disordered-to-ordered transition occurs the fastest, occurs at intermediate temperatures (540 - 555 \degree C).
Meanwhile, the transition is slower at higher and lower temperatures, a consequence of lower driving forces at high temperatures and sluggish diffusion at low temperatures (\textbf{Figure \ref{fig:Figure9}b}).
\par

From the measured TTT curve, we can estimate a critical cooling rate by drawing a linear segment from the pre-melting temperature to the peak of the nose of the TTT curve, followed by calculating the slope of the linear segment.
For Al-Ni, this critical cooling rate, R\textsubscript{c}, is -2,400 \degree C/s.
Comparing to the DSC heat flow trends in \textbf{Figure \ref{fig:Figure5}d}, the broadening of the exothermal signal, related to the change in a disordered versus ordered grain boundary, roughly occurs within the cooling rate range of -2,000 to -3,000 \degree C/s, matching the estimated critical cooling rate from the TTT curve.
Comparing this result to prior work on bulk binary Al alloys formed through powder consolidation, the Al-Y system was the only binary system shown to form amorphous defect phases at cooling rates below -1 \degree C/s, a likely consequence of Y being an efficient defect phase stabilizer and segregant species, as previously discussed \cite{LEI2023}.
Meanwhile, the other two binary systems, Al-Mg and Al-Ni, only exhibited amorphous defect phases following a water quenching procedure, with cooling rates well above -1 \degree C/s, collaborating the high critical cooling rate for Al-Ni calculated here.
Furthermore, the addition of chemical complexity has consistently been shown to reduce the critical cooling rate, with examinations of bulk ternary Al alloys (Al-Fe-Y, Al-Mg-Y, and Al-Ni-Y) revealing that each of the alloy systems was capable of maintaining amorphous defect phases at cooling rates below -1 \degree C/s \cite{LEI2021}.
Considering other amorphous defect phase containing systems outside of Al alloys, such as Cu-based alloys, reveals an identical trend, with the binary system Cu-Zr exhibiting poor amorphous defect phase formation with a critical cooling rate on the order of -$10^{6}$ \degree C/s, and the more chemical complex ternary alloy, Cu-Zr-Hf, exhibiting a critical cooling rate on par with Al-Ni, on the order of -1,000 \degree C/s \cite{GRIGORIAN2021}.
\par

Determining critical cooling rates has long been an important estimator of a metallic glass alloy's glass forming ability, which can be treated as the resistance to nucleation of crystalline phases during the solidification process \cite{davies1978}.
As such, critical cooling rates have been determined for a wide range of metallic glasses, with typical values falling within a range of $-10^2$ to $-10^8$ \degree C/s \cite{takeuchi2001}.
Of more interest to this work are the critical cooling rates for Al-based metallic glasses, which for the broad collection of critical cooling rates compiled in Ref. \cite{LIAO2015}, have values that fall within a narrow range of $-10^3$ to $-10^4$ \degree C/s.
Al-based metallic glasses are generally considered poor metallic glass formers as Al-based alloys do not tend to form as eutectics near Al-rich compositions (\textbf{Figure \ref{fig:Figure3}}).
Hence, most literature on successful Al-based metallic glasses has focused on ternary, quaternary, and quinary compositions that typically contain some elemental distribution of Al, a transition metal, and a rare earth element \cite{INOUE1998}.
Significantly, the critical cooling rate for the grain boundary confined disordered-to-ordered transition explored here in the binary Al-Ni system (R\textsubscript{c} = -2,400 \degree C/s) falls within the critical cooling rate range for more chemically complex metallic glass alloys.
We propose that this reduction in critical cooling rate for a simple binary system follows from: one, enhanced diffusion pathways along grain boundaries that are a consequence of the excess free volume characteristic of amorphous structures \cite{GRIGORIAN2021}, two, the interfacial solute segregation mechanism drives localized amorphization without requiring global chemical homogeneity, which is typically necessary in chemically complex BMGs to achieve geometric frustration \cite{miller2007}, and three, the confinement of the grains itself restricts crystallization, as BMGs have surrounding disordered material while the grain boundary confined defect phase has less flexible/deformable crystalline material nearby. 
This behavior, supported by prior experimental evidence from Ref. \cite{LEI2021}, underpins that even simple binary amorphous defect phase containing systems exhibit critical cooling rates on the same order of magnitude as chemical complex BMGs, and that the addition of chemical complexity to these Al-based amorphous defect phase forming systems will serve to further reduce the critical cooling rate for the disordered-to-ordered transition, facilitating bulk sample processing approaches.
\par

\subsection{Recovery of the Amorphous Defect Phase}

\begin{figure} [htbp]
    \centering
        \includegraphics[width=1.0\linewidth]{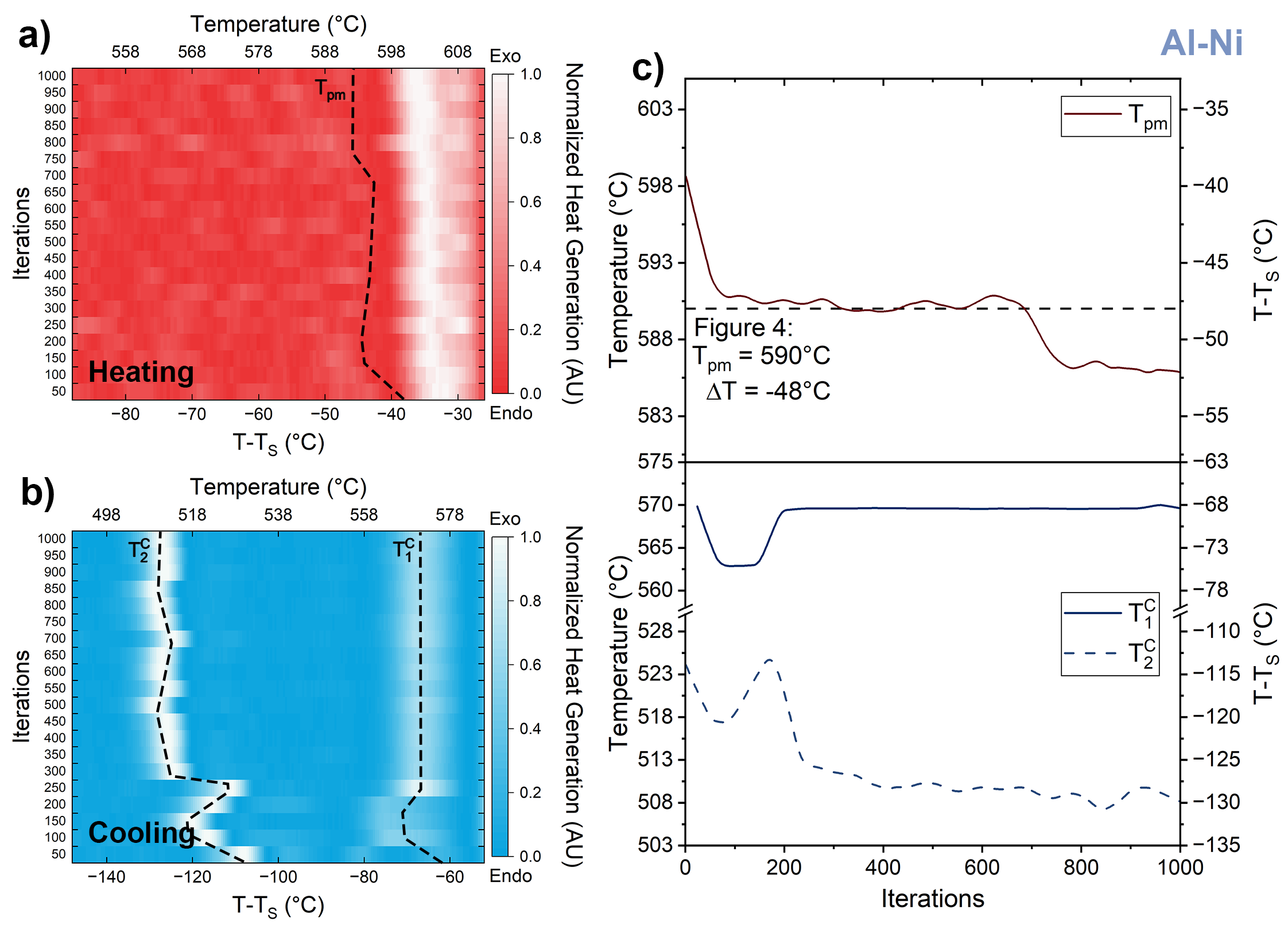}
        \caption{Normalized and background-subtracted DSC heat flow curves across 1000 repeated DSC runs (-50 to 625 \degree C at heating/cooling rates of 5,000 \degree C/s) for Al-Ni during (a) heating and (b) cooling. (c) The onset of the pre-melting regime (T\textsubscript{pm}) as a function of the number of DSC run iterations. The dashed line is the estimated onset of the pre-melting regime determined in \textbf{Figure \ref{fig:Figure5}d}. (d) Two exothermic signals during cooling ($\text{T}_1^\text{C}$ and $\text{T}_2^\text{C}$) as a function of the number of DSC run iterations.}
        \label{fig:Figure10}
\end{figure}

A critical feature of the amorphous defect phase is that the disordered grain boundary structure is the preferred structure at sub-eutectic temperatures as supported by the relative insensitivity of the pre-melting temperatures measured by DSC to the previous thermal history.
As such, the amorphous defect phase should be capable of indefinite nucleation and relaxation, as long as the dopant species remains predominately grain boundary segregated.
To explore the stability of the amorphous defect phase, we apply a repeated thermal profile (1000 runs of -50 to 620 \degree C at heating/cooling rates of 5,000 \degree C/s; \textbf{Figure \ref{fig:Figure1}e}) to Al-Ni; normalized and background-subtracted heat flow curves from heating and cooling are provided in \textbf{Figures \ref{fig:Figure10}a} and \textbf{b}, respectively.
Examination of the heating curve (\textbf{Figure \ref{fig:Figure10}a}) indicated a consistent signature across all iterations: the onset of the pre-melting regime (T\textsubscript{pm}), plotted in \textbf{Figure \ref{fig:Figure10}c} as a function of the number of iterations with an applied average smoothing filter to facilitate comprehension of the data. 
The onset of the pre-melting regime, T\textsubscript{pm}, begins at an initial value of 598 \degree C ($\mid$$\Delta$T$\mid$ $=$ 40 \degree C) and subsequently drops to nominally 590 \degree C ($\mid$$\Delta$T$\mid$ $=$ 48 \degree C) over the first 100 iterations.
Significantly, this temperature matches the onset of the pre-melting regime estimated from the Al-Ni curves at different cooling rates in \textbf{Figure \ref{fig:Figure5}d}, noted by the dashed line in \textbf{Figure \ref{fig:Figure10}c}.
After approximately 700 iterations, T\textsubscript{pm} again drops to 585.5 \degree C ($\mid$$\Delta$T$\mid$ $=$ 52.5 \degree C), remaining at this temperature through the full 1000 iterations.
\par

Examination of the cooling curves (\textbf{Figure \ref{fig:Figure10}b}) reveals a more complex series of behaviors consisting of two exothermic signatures: $\text{T}_1^\text{C}$ and $\text{T}_2^\text{C}$.
$\text{T}_1^\text{C}$ is the most stable exothermic signal, starting at nominally 570 \degree C ($\mid$$\Delta$T$\mid$ $=$ 68 \degree C) with a local minima after 100 iterations at 563 \degree C ($\mid$$\Delta$T$\mid$ $=$ 75 \degree C).
The signal then stabilizes again at 570 \degree C ($\mid$$\Delta$T$\mid$ $=$ 68 \degree C) after approximately 200 iterations and remains at this value for all subsequent iterations.
$\text{T}_2^\text{C}$, on the other hand, goes from 523 \degree C ($\mid$$\Delta$T$\mid$ $=$ 115 \degree C) to a minima of 510.5 \degree C ($\mid$$\Delta$T$\mid$ $=$ 127.5 \degree C) after 400 iterations and remains roughly at this temperature for all following iterations.
Curiously, neither of these exothermic peaks match the behavior shown in \textbf{Figure \ref{fig:Figure5}d}, which consisted of a single exothermic signal at 538 \degree C ($\mid$$\Delta$T$\mid$ $=$ 100 \degree C), a likely reflection of the quick annealing times of the first thermal profile (\textbf{Figure \ref{fig:Figure1}c}).
We hypothesize that the first exothermic signal, $\text{T}_1^\text{C}$, is related to the disordered grain boundary transition, which is broad throughout all iterations due to the high cooling rate of -5,000 \degree C/s, while the second exothermic signal, $\text{T}_2^\text{C}$, is likely related to crystallization of the Al\textsubscript{3}Ni intermetallic.
Local compositional fluctuations enable partial ordering of Ni-rich regions during cooling, which at critical undercooling thresholds can undergo rapid crystallization into Al\textsubscript{3}Ni, releasing latent heat \cite{GRAPES2017, JACKSON1999}.
Indeed, isothermal experiments on Al-Ni multilayers reveal that Al\textsubscript{3}Ni formation occurs even at heating/cooling rates exceeding 10,000 K/s, with activation energies of $\sim$150–200 kJ/mol \cite{GRAPES2017}.
Significantly, all trends, regardless of heating or cooling, show an initial significant reduction within the first 200 iterations.
This behavior suggests the presence of at least one significant microstructural change early in the thermal cycling process and necessitates additional microscopy to elucidate the origin of this behavior.
\par

\begin{figure}
    \centering
        \includegraphics[width=0.75\linewidth]{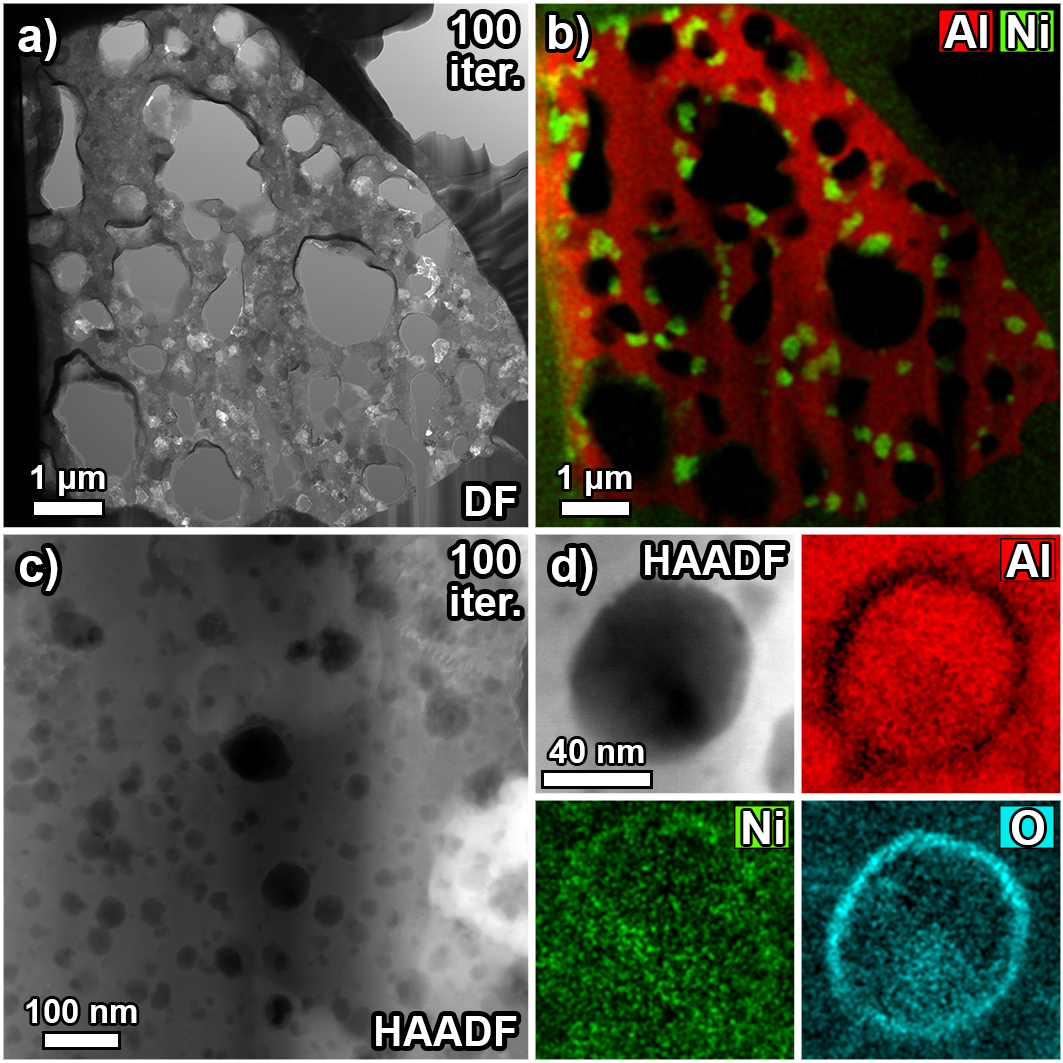}
        \caption{(a) STEM-DF micrograph of Al-Ni after 100 iterations of the thermal cycling sequence (-30 \degree C to 620 \degree C at heating/cooling rates of 5,000 \degree C/s). (b) Combined STEM-EDS map of Al (red) and Ni (green). (c) Higher magnification STEM-HAADF micrograph of Al-Ni after 100 iterations of the thermal cycling sequence demonstrating void formation. (d) STEM-HAADF micrograph of a single void with separate STEM-EDS maps of Al (red), Ni, (green), and O (cyan).} 
        \label{fig:Figure11}
\end{figure}

To capture the microstructural transitions following multiple iterations of the thermal cycling sequence, two separate Al-Ni samples were subjected to the thermal profile in \textbf{Figure \ref{fig:Figure1}e} with increasing numbers of iterations: 100 and 600 iterations.
After confirming that each sample exhibited similar trends in the heating curves through number of iterations, TEM lamellas were prepared from each condition by utilizing a FIB/SEM to lift out individual powders, which were subsequently mounted onto Cu Omniprobe grids and thinned to electron transparency.
A STEM-DF micrograph of Al-Ni after 100 iterations shows some increase in the average grain size (\textbf{Figure \ref{fig:Figure11}a}) and little change in the Ni distribution from the as-milled (\textbf{Figure \ref{fig:Figure11}b}).
The most striking feature after 100 iterations is the formation of pores, which are evident at large length-scales (500 nm - 1 $\mu$m in \textbf{Figure \ref{fig:Figure11}a}) and small spherical inclusions at the nano-scale regime (\textbf{Figure \ref{fig:Figure11}c}), and are not present in the as-milled powders (\textbf{Figure \ref{fig:Figure6}}).
STEM-EDS maps of oxygen across these spherical inclusions (\textbf{Figure \ref{fig:Figure11}d}) indicates segregation of O towards the walls of the inclusions, suggesting that these features are not nano-scale oxides as frequently reported in the literature \cite{yan2018, horn2022}, and are likely nano-scale voids that serve as precursors to pores.
Significantly, continuing the thermal cycling sequence to 600 iterations shows little difference in the microstructure compared to the 100 iteration microstructure, with similar grain sizes and pore formation, in addition to the expected nano-scale void formation (\textbf{Supplemental Figure 2}).
\par

The presence of a broad distribution of voids suggests that voids are constantly nucleating and coalescing during each iteration of the thermal profile, which would thus indicate the presence of a constant source of excess vacancies.
Early work exploring the role of quenching in pure Al has shown that significant vacancy populations are generated at high temperatures, which leads to an excess vacancy population at lower temperatures following rapid quenching \cite{Khludkova1965, Bradshaw1957}.
An upper-limit estimate of the vacancies generated during quenching can be determined by the following expression:
\begin{equation}
    C_v = \text{exp}\left(-\frac{Q_f}{kT}\right)(1-ZC_s)
\end{equation}
where $Q_f$ is the vacancy formation energy, $k$ is Boltzmann's constant, $T$ is the heat treatment temperature, and $(1-ZC_s)$ is the correction term for a binary alloy, with $Z$ the coordination number and $C_s$ the solute concentration \cite{WERT1980}.
Using 0.7 eV for $Q_f$ \cite{HUANG2023} and an annealing temperature of 620 \degree C, we calculated an upper-limit vacancy concentration of $1\times10^{-4}$ V/Al for a single iteration, a relatively large value due to the high annealing temperature.
Furthermore, we note that, if this process is truly dominated by vacancies formed during quenching, varying the quench rate should result in a change in vacancy migration, with increasing quench rates leading to reduced vacancy diffusion towards sinks (such as grain boundaries and surfaces) and resulting in larger and faster pore formation in the bulk.
We confirm this behavior by annealing Al-Ni-Y powders with a quench rate of -10,000 \degree C/s (\textbf{Supplemental Figure 3}), which after only a single iteration exhibited significant pore formation, as opposed to the Al-Y sample in \textbf{Figure \ref{fig:Figure7}}, which exhibited few obvious pores or voids after a slower quench rate of -3,000 \degree C/s.
Thus, we propose the following mechanism is responsible for the formation of the pores in each sample condition: during quenching, excess vacancies are kinetically frozen in and the subsequent annealing at higher temperatures increases vacancy mobility, enhancing vacancy-vacancy interactions and forming voids, which continue to coalesce, resulting in macroscale pores.
\par

Identification of the significant vacancy generation during quenching highlights an important question: namely, does the production of vacancies at such high rates induce any variations in the observed DSC signals?
It is important to note that the only changes observed in the DSC signals were shifts in peaks present following the first iteration, with no additional peaks or features appearing and no features disappearing upon subsequent heating and cooling.
This suggests that the presence of vacancies has little effect on the actual amorphous defect phase nucleation process despite the high rate of vacancy generation.
While this behavior could be a consequence of minimal excess vacancy diffusion towards grain boundaries, it is unlikely given the high vacancy mobility in Al at temperatures above 450 \degree C and the short diffusion lengths in nanocrystalline/ultrafine-grained microstructures \cite{SWALIN1967}.
However, excess vacancies could still have some influence on curve shape or critical temperatures.
Prior work with Al alloys has shown that excessive vacancy content can indeed drive some changes in DSC signals, including:
\begin{enumerate}
    \item \textbf{Exothermal signatures} - Quenched-in vacancies enhance solute diffusion, which in turn facilitates vacancy cluster and/or precipitate formation during subsequent annealing. 
    This behavior could manifest as additional exothermal peaks during heating or as a shift in precipitation kinetics \cite{Banhart2020, milkereit2019}. 
    While no additional exothermal peaks are present upon subsequent iterations, there is a shift in the onset of the pre-melting regime towards lower values before 200 iterations, which could be a consequence of vacancy-enhanced solute diffusion.
    \item \textbf{Endothermic signatures} - If metastable phases are formed during prior iterations due to vacancy-enhanced solute diffusion, then dissolution of these phases would be captured in the DSC heat flows, appearing as additional endothermic signals during heating \cite{HARDY1954, Lotter2018}.
    As no additional endothermic signatures appear in any of our DSC heat flow curves during subsequent iterations, it is unlikely that this mechanism is relevant.
\end{enumerate}
The DSC signals in \textbf{Figure \ref{fig:Figure10}} indicate little change above 200 iterations, in both cooling and heating curves, with only a $< 5$ \degree C change observed in the heating curve.
As vacancies are constantly generated through each quenching process, the consistent behavior above 200 iterations suggests that the influence of vacancies has little long-term influence on the thermal signatures.
The strongest transition occurs before 200 iterations, with a reduction in the onset of the pre-melting regime and in both observed exothermal signals during cooling.
\par

\begin{figure}
    \centering
        \includegraphics[width=0.75\linewidth]{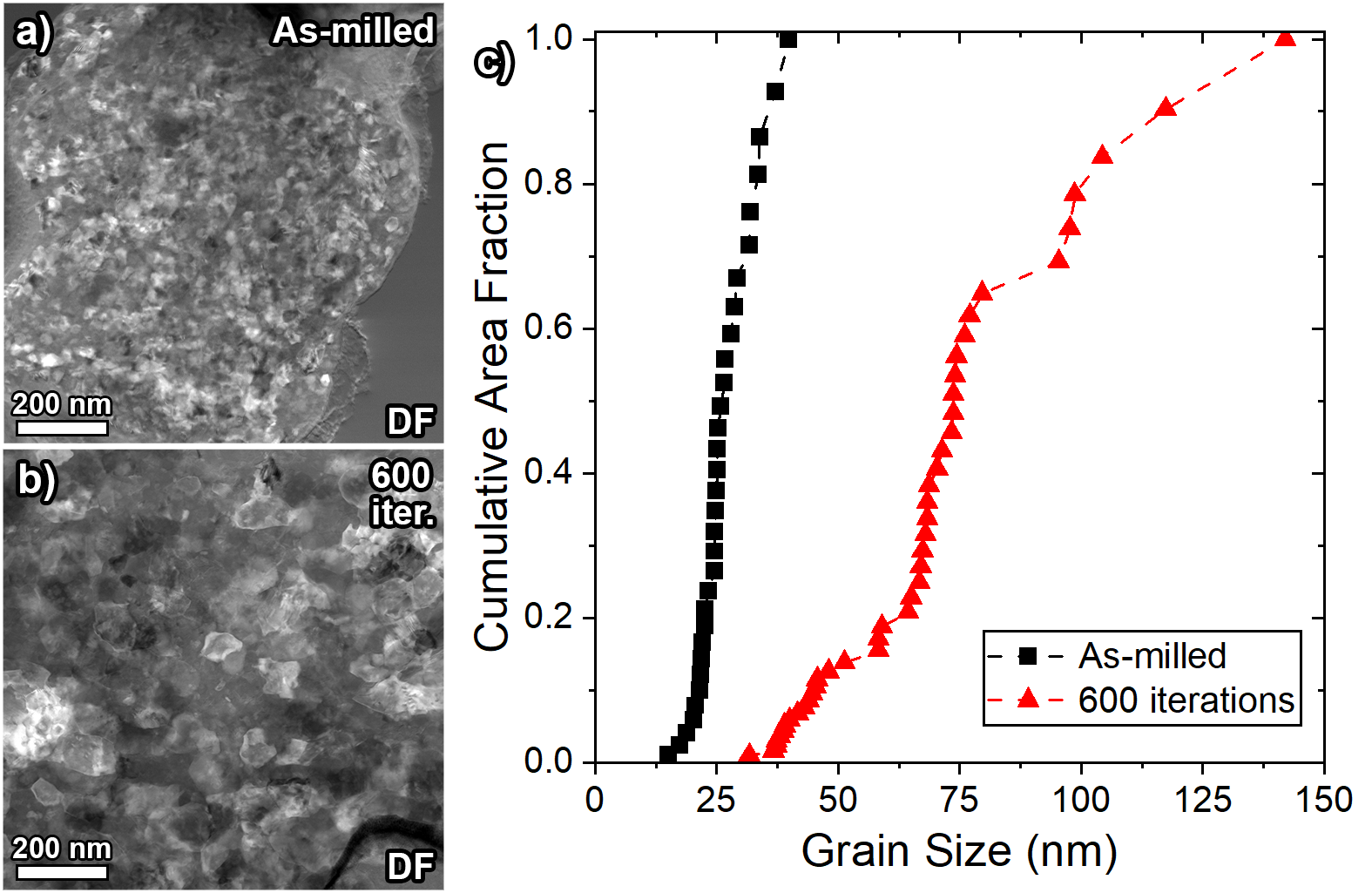}
        \caption{(a) STEM-DF micrograph of as-milled Al-Ni. (a) STEM-DF micrograph of Al-Ni after 600 iterations of the thermal cycling sequence (-30 \degree C to 620 \degree C at heating/cooling rates of 5,000 \degree C/s). (c) Cumulative grain size distributions for the as-milled Al-Ni (black) and Al-Ni after 600 iterations (red).} 
        \label{fig:Figure12}
\end{figure}

We propose that the shifts in the exothermal signals upon cooling and the pre-melting signature upon heating are a consequence of vacancy-enhanced diffusion of solute atoms towards the grain boundaries.
At a fundamental level, increasing solute concentrations at the grain boundaries directly reduces the interfacial energy ($\gamma_{GB}$) by counteracting the excess interfacial free energy, as described in the fundamental works of Cahn \cite{cahn1979} and Weissmüller \cite{weissmuller1993}, and has been demonstrated in a number of stabilized nanocrystalline systems \cite{chookajorn2012, Cunningham2022, darling2013, devaraj2019}.
Furthermore, increasing grain boundary solute also drives a reduction in the volumetric free energy penalty for forming an undercooled liquid ($\Delta G$\textsubscript{amorphous}).
High solute configurations increase configurational entropy, offsetting the entropic penalty of undercooling, while simultaneously increasing the number of solute-solute interactions, which lowers the enthalpy difference between the crystalline and amorphous states \cite{ZHOU2016}.
Thus, as solute concentrations increase at the grain boundaries, the propensity for amorphous defect phase formation also increases, lowering the onset of the pre-melting regime.
An important consequence of this increased propensity for amorphous defect phase formation can be observed in the microstructure, which, despite repeated annealing sequences at homologous temperatures above 0.9T\textsubscript{M}, was able to maintain a fine-grained structure (\textbf{Figure \ref{fig:Figure12}}).
Indeed, grain size distributions (\textbf{Figure \ref{fig:Figure12}c}) for the as-milled and post 600 iterations microstructure indicated only a minor increase, going from average grain sizes of nominally 25 nm to approximately 75 nm, illustrating the remarkable ability of the amorphous defect phase to withstand repeated annealing sequences and stabilize the microstructure against such temperature extremes.
\par

\section{Conclusions}
\label{sec:conclusions}

In this work, we utilized the fast heating and cooling rates from ultrafast DSC to probe the grain boundary confined amorphous transition in four nanocrystalline systems: Al-Ni, Al-Y, Al-Ni-Y, and Al-Mg-Y.
Together with correlative TEM, we revealed the following mechanisms:
\begin{itemize}
    \item All alloy systems exhibited a distinct pre-melting signature upon heating. 
    The range of this pre-melting feature, $\Delta$T, or the onset of the pre-melting regime relative to the eutectic, serves as a useful experimental metric for amorphous defect phase forming ability, with alloy systems exhibiting a higher $\mid$$\Delta$T$\mid$ demonstrating a lower barrier for amorphous defect phase formation.
    Significantly, this pre-melting transition is largely insensitive to prior thermal history.
    \item Increases in the cooling rate across all alloy systems exhibited a clear transition from grain boundary ordering to kinetic freezing of the amorphous grain boundary phase with higher cooling rates. 
    Further variations in exothermal peak numbers can be attributed to additional microstructural complexity from intermetallic formation.
    \item Increasing the chemical complexity, through addition of a second solute species, promoted amorphous defect phase formation.
    \item The kinetic behavior of Al-Ni was captured through a series of sequential isothermal measurements, resulting in a TTT curve for the grain boundary confined phase transition. 
    From this TTT curve, a critical cooling rate of -2,400 \degree C/s was determined for the grain boundary confined disordered-to-ordered transition in Al-Ni.
    \item Recovery of the amorphous defect phase was explored in Al-Ni through a series of 1,000 annealing runs to 620 \degree C. 
    Examination of the DSC heat flow curves showed remarkable stability through heating and cooling, with no change in the number of features present through increasing iterations.
    However, the high quench rates drove significant vacancy generation, which in turn promoted diffusion of the solute species towards the grain boundaries, shifting the onset of the pre-melting regime.
    \item Nanocrystallinity of the powders was maintained through repeated annealing at homologous temperatures above 0.9T\textsubscript{M}, demonstrating the remarkable thermal stability of these alloys.
\end{itemize}
Overall, this study provides comprehensive insights into the mechanisms governing grain boundary confined amorphous transitions in nanocrystalline alloys.
Further developing our understanding of such transitions can be leveraged, through a \textit{defects by design} strategy, as a pathway to create novel alloys with enhanced material properties that push the boundaries of high temperature stability.
\par

\section*{CRediT authorship contribution statement}
\label{sec:credit_statement}

\textbf{W.S. Cunningham:} Writing – review \& editing, Writing – original draft, Project administration, Visualization, Methodology, Funding acquisition, Investigation, Formal analysis, Conceptualization, Data curation. 
\textbf{T. Lei:} Writing – review \& editing, Visualization, Investigation, Formal analysis.
\textbf{H.C. Howard:} Writing – review \& editing, Methodology, Investigation, Formal analysis.  
\textbf{T.J. Rupert:} Writing – review \& editing, Methodology, Conceptualization. 
\textbf{D.S. Gianola:} Writing – review \& editing, Visualization, Project administration, Methodology, Funding acquisition, Formal analysis, Conceptualization.

\section*{Declaration of Competing Interest}
\label{sec:competing_interests}

The authors declare that they have no known competing financial interests or personal relationships that could have appeared to influence the work reported in this paper.

\section*{Acknowledgments}
\label{sec:acknowledgments}

This material is based upon work supported by the National Science Foundation MPS-Ascend Postdoctoral Research Fellowship under Grant No. 2316692.
The research reported here made use of the shared facilities of the Materials Research Science and Engineering Center (MRSEC) at UC Santa Barbara: NSF DMR–2308708. 
The UC Santa Barbara MRSEC is a member of the Materials Research Facilities Network (www.mrfn.org).
HCH acknowledges support from the National Science Foundation Graduate
Research Fellowship Program under Grant No. 2139319.
Any opinions, findings, and conclusions or recommendations expressed in this material are those of the author(s) and do not necessarily reflect the views of the National Science Foundation.
TJR knowledges support from the U.S. Department of Energy, Office of Science, Basic Energy Sciences under Award No. DE-SC0025195.
DSG and WSC acknowledge additional support from the Army Research Laboratory under Cooperative Agreement Number W911NF-22-2-0121. 
The views and conclusions contained in this document are those of the authors and should not be interpreted as representing the official policies, either expressed or implied, of the Army Research Laboratory or the U.S. Government. 
The U.S. Government is authorized to reproduce and distribute reprints for Government purposes notwithstanding any copyright notation herein.
\par

\bibliographystyle{elsarticle-num} 
\bibliography{cas-refs}

\begin{thebibliography}{100}
\expandafter\ifx\csname url\endcsname\relax
  \def\url#1{\texttt{#1}}\fi
\expandafter\ifx\csname urlprefix\endcsname\relax\def\urlprefix{URL }\fi
\expandafter\ifx\csname href\endcsname\relax
  \def\href#1#2{#2} \def\path#1{#1}\fi

\bibitem{Suryanarayana1995}
C.~Suryanarayana, Nanocrystalline materials, International Materials Reviews 40~(2) (1995) 41--64.
\newblock \href {https://doi.org/10.1179/imr.1995.40.2.41} {\path{doi:10.1179/imr.1995.40.2.41}}.

\bibitem{KUMAR2003}
K.~Kumar, H.~{Van Swygenhoven}, S.~Suresh, Mechanical behavior of nanocrystalline metals and alloys, Acta Materialia 51~(19) (2003) 5743--5774.
\newblock \href {https://doi.org/10.1016/j.actamat.2003.08.032} {\path{doi:10.1016/j.actamat.2003.08.032}}.

\bibitem{MEYERS2006}
M.~Meyers, A.~Mishra, D.~Benson, Mechanical properties of nanocrystalline materials, Progress in Materials Science 51~(4) (2006) 427--556.
\newblock \href {https://doi.org/10.1016/j.pmatsci.2005.08.003} {\path{doi:10.1016/j.pmatsci.2005.08.003}}.

\bibitem{Heckman2018}
N.~M. Heckman, S.~M. Foiles, C.~J. O{'}Brien, M.~Chandross, C.~M. Barr, N.~Argibay, K.~Hattar, P.~Lu, D.~P. Adams, B.~L. Boyce, New nanoscale toughening mechanisms mitigate embrittlement in binary nanocrystalline alloys, Nanoscale 10 (2018) 21231--21243.
\newblock \href {https://doi.org/10.1039/C8NR06419A} {\path{doi:10.1039/C8NR06419A}}.

\bibitem{ZHANG2006}
Y.~Zhang, Z.~Han, K.~Wang, K.~Lu, Friction and wear behaviors of nanocrystalline surface layer of pure copper, Wear 260~(9) (2006) 942--948.
\newblock \href {https://doi.org/10.1016/j.wear.2005.06.010} {\path{doi:10.1016/j.wear.2005.06.010}}.

\bibitem{LIU2010}
L.~Liu, Y.~Li, F.~Wang, Electrochemical corrosion behavior of nanocrystalline materials - a review, Journal of Materials Science \& Technology 26~(1) (2010) 1--14.
\newblock \href {https://doi.org/10.1016/S1005-0302(10)60001-1} {\path{doi:10.1016/S1005-0302(10)60001-1}}.

\bibitem{BEYERLEIN2013}
I.~Beyerlein, A.~Caro, M.~Demkowicz, N.~Mara, A.~Misra, B.~Uberuaga, Radiation damage tolerant nanomaterials, Materials Today 16~(11) (2013) 443--449.
\newblock \href {https://doi.org/10.1016/j.mattod.2013.10.019} {\path{doi:10.1016/j.mattod.2013.10.019}}.

\bibitem{CHENG2016}
G.~Cheng, W.~Xu, Y.~Wang, A.~Misra, Y.~Zhu, Grain size effect on radiation tolerance of nanocrystalline mo, Scripta Materialia 123 (2016) 90--94.
\newblock \href {https://doi.org/10.1016/j.scriptamat.2016.06.007} {\path{doi:10.1016/j.scriptamat.2016.06.007}}.

\bibitem{Korte-Kerzel2022}
S.~Korte-Kerzel, T.~Hickel, L.~Huber, D.~Raabe, S.~Sandlöbes-Haut, M.~Todorova, J.~Neugebauer, Defect phases – thermodynamics and impact on material properties, International Materials Reviews 67~(1) (2022) 89--117.
\newblock \href {https://doi.org/10.1080/09506608.2021.1930734} {\path{doi:10.1080/09506608.2021.1930734}}.

\bibitem{TEHRANCHI2024}
A.~Tehranchi, S.~Zhang, A.~Zendegani, C.~Scheu, T.~Hickel, J.~Neugebauer, Metastable defect phase diagrams as roadmap to tailor chemically driven defect formation, Acta Materialia 277 (2024) 120145.
\newblock \href {https://doi.org/10.1016/j.actamat.2024.120145} {\path{doi:10.1016/j.actamat.2024.120145}}.

\bibitem{MATSON2024}
T.~P. Matson, C.~A. Schuh, Phase and defect diagrams based on spectral grain boundary segregation: A regular solution approach, Acta Materialia 265 (2024) 119584.
\newblock \href {https://doi.org/10.1016/j.actamat.2023.119584} {\path{doi:10.1016/j.actamat.2023.119584}}.

\bibitem{SHI2011}
X.~Shi, J.~Luo, Developing grain boundary diagrams as a materials science tool: A case study of nickel-doped molybdenum, Phys. Rev. B 84 (2011) 014105.
\newblock \href {https://doi.org/10.1103/PhysRevB.84.014105} {\path{doi:10.1103/PhysRevB.84.014105}}.

\bibitem{DILLON2007}
S.~J. Dillon, M.~Tang, W.~C. Carter, M.~P. Harmer, Complexion: A new concept for kinetic engineering in materials science, Acta Materialia 55~(18) (2007) 6208--6218.
\newblock \href {https://doi.org/10.1016/j.actamat.2007.07.029} {\path{doi:10.1016/j.actamat.2007.07.029}}.

\bibitem{KIRCHHEIM2002}
R.~Kirchheim, Grain coarsening inhibited by solute segregation, Acta Materialia 50~(2) (2002) 413--419.
\newblock \href {https://doi.org/10.1016/S1359-6454(01)00338-X} {\path{doi:10.1016/S1359-6454(01)00338-X}}.

\bibitem{gibbs1878}
J.~W. Gibbs, On the equilibrium of heterogeneous substances, American Journal of Science 3~(96) (1878) 441--458.

\bibitem{Frolov2015}
T.~Frolov, Y.~Mishin, {Phases, phase equilibria, and phase rules in low-dimensional systems}, The Journal of Chemical Physics 143~(4) (2015) 044706.
\newblock \href {https://doi.org/10.1063/1.4927414} {\path{doi:10.1063/1.4927414}}.

\bibitem{CANTWELL2020}
P.~R. Cantwell, T.~Frolov, T.~J. Rupert, A.~R. Krause, C.~J. Marvel, G.~S. Rohrer, J.~M. Rickman, M.~P. Harmer, Grain boundary complexion transitions, Annual Review of Materials Research 50~(Volume 50, 2020) (2020) 465--492.
\newblock \href {https://doi.org/10.1146/annurev-matsci-081619-114055} {\path{doi:10.1146/annurev-matsci-081619-114055}}.

\bibitem{CANTWELL2014}
P.~R. Cantwell, M.~Tang, S.~J. Dillon, J.~Luo, G.~S. Rohrer, M.~P. Harmer, Grain boundary complexions, Acta Materialia 62 (2014) 1--48.
\newblock \href {https://doi.org/10.1016/j.actamat.2013.07.037} {\path{doi:10.1016/j.actamat.2013.07.037}}.

\bibitem{HOWARD2025}
H.~Howard, W.~Cunningham, A.~Genc, B.~Rhodes, B.~Merle, T.~Rupert, D.~Gianola, Chemically ordered dislocation defect phases as a new strengthening pathway in ni–al alloys, Acta Materialia 289 (2025) 120887.
\newblock \href {https://doi.org/10.1016/j.actamat.2025.120887} {\path{doi:10.1016/j.actamat.2025.120887}}.

\bibitem{TURLO2020}
V.~Turlo, T.~J. Rupert, Prediction of a wide variety of linear complexions in face centered cubic alloys, Acta Materialia 185 (2020) 129--141.
\newblock \href {https://doi.org/10.1016/j.actamat.2019.11.069} {\path{doi:10.1016/j.actamat.2019.11.069}}.

\bibitem{KUZMINA2015}
M.~Kuzmina, M.~Herbig, D.~Ponge, S.~Sandlöbes, D.~Raabe, Linear complexions: Confined chemical and structural states at dislocations, Science 349~(6252) (2015) 1080--1083.
\newblock \href {https://doi.org/10.1126/science.aab2633} {\path{doi:10.1126/science.aab2633}}.

\bibitem{SINGH2023}
D.~Singh, V.~Turlo, D.~S. Gianola, T.~J. Rupert, Linear complexions directly modify dislocation motion in face-centered cubic alloys, Materials Science and Engineering: A 870 (2023) 144875.
\newblock \href {https://doi.org/10.1016/j.msea.2023.144875} {\path{doi:10.1016/j.msea.2023.144875}}.

\bibitem{smith2018}
T.~M. Smith, B.~D. Esser, B.~Good, M.~Hooshmand, G.~B. Viswanathan, C.~M.~F. Rae, M.~Ghazisaeidi, D.~W. McComb, M.~J. Mills, Segregation and phase transformations along superlattice intrinsic stacking faults in ni-based superalloys, Metallurgical and Materials Transactions A 49 (2018) 4186--4198.

\bibitem{titus2016}
M.~S. Titus, R.~K. Rhein, P.~B. Wells, P.~C. Dodge, G.~B. Viswanathan, M.~J. Mills, A.~Van~der Ven, T.~M. Pollock, Solute segregation and deviation from bulk thermodynamics at nanoscale crystalline defects, Science advances 2~(12) (2016) e1601796.

\bibitem{Frolov2015PhysRevB}
T.~Frolov, M.~Asta, Y.~Mishin, Segregation-induced phase transformations in grain boundaries, Phys. Rev. B 92 (2015) 020103.
\newblock \href {https://doi.org/10.1103/PhysRevB.92.020103} {\path{doi:10.1103/PhysRevB.92.020103}}.

\bibitem{Frolov2013}
T.~Frolov, D.~L. Olmsted, M.~Asta, Y.~Mishin, Structural phase transformations in metallic grain boundaries, Nature communications 4~(1) (2013) 1899.

\bibitem{MISHIN2010}
Y.~Mishin, M.~Asta, J.~Li, Atomistic modeling of interfaces and their impact on microstructure and properties, Acta Materialia 58~(4) (2010) 1117--1151.
\newblock \href {https://doi.org/10.1016/j.actamat.2009.10.049} {\path{doi:10.1016/j.actamat.2009.10.049}}.

\bibitem{SCHULER2017}
J.~D. Schuler, T.~J. Rupert, Materials selection rules for amorphous complexion formation in binary metallic alloys, Acta Materialia 140 (2017) 196--205.
\newblock \href {https://doi.org/10.1016/j.actamat.2017.08.042} {\path{doi:10.1016/j.actamat.2017.08.042}}.

\bibitem{LEI2023}
T.~Lei, E.~C. Hessong, D.~S. Gianola, T.~J. Rupert, Binary nanocrystalline alloys with strong glass forming interfacial regions: Complexion stability, segregation competition, and diffusion pathways, Materials Characterization 206 (2023) 113415.
\newblock \href {https://doi.org/10.1016/j.matchar.2023.113415} {\path{doi:10.1016/j.matchar.2023.113415}}.

\bibitem{LUO2005}
J.~Luo, V.~K. Gupta, D.~H. Yoon, I.~Meyer, H.~M., Segregation-induced grain boundary premelting in nickel-doped tungsten, Applied Physics Letters 87~(23) (2005) 231902.
\newblock \href {https://doi.org/10.1063/1.2138796} {\path{doi:10.1063/1.2138796}}.

\bibitem{LUO2008}
J.~Luo, Liquid-like interface complexion: From activated sintering to grain boundary diagrams, Current Opinion in Solid State and Materials Science 12~(5) (2008) 81--88.
\newblock \href {https://doi.org/10.1016/j.cossms.2008.12.001} {\path{doi:10.1016/j.cossms.2008.12.001}}.

\bibitem{LI2019}
F.~Li, T.~Liu, J.~Zhang, S.~Shuang, Q.~Wang, A.~Wang, J.~Wang, Y.~Yang, Amorphous–nanocrystalline alloys: fabrication, properties, and applications, Materials Today Advances 4 (2019) 100027.
\newblock \href {https://doi.org/10.1016/j.mtadv.2019.100027} {\path{doi:10.1016/j.mtadv.2019.100027}}.

\bibitem{williams2009}
P.~Williams, Y.~Mishin, Thermodynamics of grain boundary premelting in alloys. ii. atomistic simulation, Acta Materialia 57~(13) (2009) 3786--3794.

\bibitem{KAPLAN2006}
W.~D. Kaplan, Y.~Kauffmann, Structural order in liquids induced by interfaces with crystals, Annual Review of Materials Research 36~(Volume 36, 2006) (2006) 1--48.
\newblock \href {https://doi.org/10.1146/annurev.matsci.36.020105.104035} {\path{doi:10.1146/annurev.matsci.36.020105.104035}}.

\bibitem{MELLENTHIN2008}
J.~Mellenthin, A.~Karma, M.~Plapp, Phase-field crystal study of grain-boundary premelting, Phys. Rev. B 78 (2008) 184110.
\newblock \href {https://doi.org/10.1103/PhysRevB.78.184110} {\path{doi:10.1103/PhysRevB.78.184110}}.

\bibitem{TANG2006}
M.~Tang, W.~C. Carter, R.~M. Cannon, Grain boundary transitions in binary alloys, Phys. Rev. Lett. 97 (2006) 075502.
\newblock \href {https://doi.org/10.1103/PhysRevLett.97.075502} {\path{doi:10.1103/PhysRevLett.97.075502}}.

\bibitem{straumal2004}
B.~Straumal, B.~Baretzky, Grain boundary phase transitions and their influence on properties of polycrystals, Interface Science 12 (2004) 147--155.

\bibitem{kikuchi1980}
R.~Kikuchi, J.~W. Cahn, Grain-boundary melting transition in a two-dimensional lattice-gas model, Physical Review B 21~(5) (1980) 1893.

\bibitem{raabe2014}
D.~Raabe, M.~Herbig, S.~Sandl{\"o}bes, Y.~Li, D.~Tytko, M.~Kuzmina, D.~Ponge, P.-P. Choi, Grain boundary segregation engineering in metallic alloys: A pathway to the design of interfaces, Current Opinion in Solid State and Materials Science 18~(4) (2014) 253--261.

\bibitem{schuler2018}
J.~D. Schuler, O.~K. Donaldson, T.~J. Rupert, Amorphous complexions enable a new region of high temperature stability in nanocrystalline ni-w, Scripta Materialia 154 (2018) 49--53.

\bibitem{BALBUS2021}
G.~H. Balbus, J.~Kappacher, D.~J. Sprouster, F.~Wang, J.~Shin, Y.~M. Eggeler, T.~J. Rupert, J.~R. Trelewicz, D.~Kiener, V.~Maier-Kiener, D.~S. Gianola, Disordered interfaces enable high temperature thermal stability and strength in a nanocrystalline aluminum alloy, Acta Materialia 215 (2021) 116973.
\newblock \href {https://doi.org/10.1016/j.actamat.2021.116973} {\path{doi:10.1016/j.actamat.2021.116973}}.

\bibitem{SHIN2022}
J.~Shin, F.~Wang, G.~H. Balbus, T.~Lei, T.~J. Rupert, D.~S. Gianola, Optimizing thermal stability and mechanical behavior in segregation-engineered nanocrystalline al--ni--ce alloys: A combinatorial study, Journal of Materials Research 37~(18) (2022) 3083--3098.

\bibitem{LEI2021}
T.~Lei, J.~Shin, D.~S. Gianola, T.~J. Rupert, Bulk nanocrystalline al alloys with hierarchical reinforcement structures via grain boundary segregation and complexion formation, Acta Materialia 221 (2021) 117394.
\newblock \href {https://doi.org/10.1016/j.actamat.2021.117394} {\path{doi:10.1016/j.actamat.2021.117394}}.

\bibitem{LEI2022}
T.~Lei, M.~Xu, J.~Shin, D.~S. Gianola, T.~J. Rupert, Growth and structural transitions of core-shell nanorods in nanocrystalline al-ni-y, Scripta Materialia 211 (2022) 114502.
\newblock \href {https://doi.org/10.1016/j.scriptamat.2022.114502} {\path{doi:10.1016/j.scriptamat.2022.114502}}.

\bibitem{grigorian2019}
C.~M. Grigorian, T.~J. Rupert, Thick amorphous complexion formation and extreme thermal stability in ternary nanocrystalline cu-zr-hf alloys, Acta Materialia 179 (2019) 172--182.

\bibitem{GRIGORIAN2021}
C.~M. Grigorian, T.~J. Rupert, Critical cooling rates for amorphous-to-ordered complexion transitions in cu-rich nanocrystalline alloys, Acta Materialia 206 (2021) 116650.
\newblock \href {https://doi.org/10.1016/j.actamat.2021.116650} {\path{doi:10.1016/j.actamat.2021.116650}}.

\bibitem{takeuchi2001}
A.~Takeuchi, A.~Inoue, Quantitative evaluation of critical cooling rate for metallic glasses, Materials Science and Engineering: A 304 (2001) 446--451.

\bibitem{grigorian2022}
C.~M. Grigorian, T.~J. Rupert, Multi-principal element grain boundaries: Stabilizing nanocrystalline grains with thick amorphous complexions, Journal of Materials Research (2022) 1--13.

\bibitem{khalajhedayati2016}
A.~Khalajhedayati, Z.~Pan, T.~J. Rupert, Manipulating the interfacial structure of nanomaterials to achieve a unique combination of strength and ductility, Nature communications 7~(1) (2016) 10802.

\bibitem{CUNNINGHAM2023}
W.~S. Cunningham, J.~Shin, T.~Lei, T.~J. Rupert, D.~S. Gianola, High-throughput assessment of the microstructural stability of segregation-engineered nanocrystalline al-ni-y alloys, Materialia 32 (2023) 101940.
\newblock \href {https://doi.org/10.1016/j.mtla.2023.101940} {\path{doi:10.1016/j.mtla.2023.101940}}.

\bibitem{YAN2023}
Z.~Yan, Z.~Liu, B.~Yao, Q.~An, R.~Zhang, S.~Zheng, Effect of amorphous complexions on plastic deformation of nanolayered composites, Scripta Materialia 231 (2023) 115470.
\newblock \href {https://doi.org/10.1016/j.scriptamat.2023.115470} {\path{doi:10.1016/j.scriptamat.2023.115470}}.

\bibitem{SCHULER2020}
J.~D. Schuler, C.~M. Grigorian, C.~M. Barr, B.~L. Boyce, K.~Hattar, T.~J. Rupert, Amorphous intergranular films mitigate radiation damage in nanocrystalline cu-zr, Acta Materialia 186 (2020) 341--354.
\newblock \href {https://doi.org/10.1016/j.actamat.2019.12.048} {\path{doi:10.1016/j.actamat.2019.12.048}}.

\bibitem{ludy2016}
J.~E. Ludy, T.~J. Rupert, Amorphous intergranular films act as ultra-efficient point defect sinks during collision cascades, Scripta Materialia 110 (2016) 37--40.

\bibitem{aksoy2024}
D.~Aksoy, P.~Cao, J.~R. Trelewicz, J.~P. Wharry, T.~J. Rupert, Enhanced radiation damage tolerance of amorphous interphase and grain boundary complexions in cu-ta, JOM (2024) 1--14.

\bibitem{Wang2021}
Y.~Wang, M.~Ostrowska, G.~Cacciamani, Thermodynamic modeling of selected ternary systems containing y and calphad simulation of conicraly metallic coatings, Calphad 72 (2021) 102214.
\newblock \href {https://doi.org/10.1016/j.calphad.2020.102214} {\path{doi:10.1016/j.calphad.2020.102214}}.

\bibitem{Chen2024}
Y.~Chen, J.~Wang, W.~Zheng, Q.~Li, M.~Yu, T.~Ying, X.~Zeng, Calphad-guided design of mg-y-al alloy with improved strength and ductility via regulating the lpso phase, Acta Materialia 263 (2024) 119521.
\newblock \href {https://doi.org/10.1016/j.actamat.2023.119521} {\path{doi:10.1016/j.actamat.2023.119521}}.

\bibitem{murdoch2013}
H.~A. Murdoch, C.~A. Schuh, Estimation of grain boundary segregation enthalpy and its role in stable nanocrystalline alloy design, Journal of Materials Research 28~(16) (2013) 2154--2163.

\bibitem{INOUE2000}
A.~Inoue, Stabilization of metallic supercooled liquid and bulk amorphous alloys, Acta Materialia 48~(1) (2000) 279--306.
\newblock \href {https://doi.org/10.1016/S1359-6454(99)00300-6} {\path{doi:10.1016/S1359-6454(99)00300-6}}.

\bibitem{atwater2012}
M.~A. Atwater, K.~A. Darling, A visual library of stability in binary metallic systems: the stabilization of nanocrystalline grain size by solute addition: part 1, Army Research Lab Aberdeen Proving Ground MD Weapons and Materials Research Directorate (2012).

\bibitem{EGAMI1984}
T.~Egami, Y.~Waseda, Atomic size effect on the formability of metallic glasses, Journal of Non-Crystalline Solids 64~(1) (1984) 113--134.
\newblock \href {https://doi.org/10.1016/0022-3093(84)90210-2} {\path{doi:10.1016/0022-3093(84)90210-2}}.

\bibitem{CHENG2011}
Y.~Cheng, E.~Ma, Atomic-level structure and structure–property relationship in metallic glasses, Progress in Materials Science 56~(4) (2011) 379--473.
\newblock \href {https://doi.org/10.1016/j.pmatsci.2010.12.002} {\path{doi:10.1016/j.pmatsci.2010.12.002}}.

\bibitem{okamoto2004}
H.~Okamoto, Al-ni (aluminum-nickel), Journal of Phase Equilibria and diffusion 25~(4) (2004) 394.

\bibitem{okamoto1998}
H.~Okamoto, Al-mg (aluminum-magnesium), Journal of Phase Equilibria and Diffusion 19~(6) (1998) 598.

\bibitem{LIU2006}
S.~Liu, Y.~Du, H.~Xu, C.~He, J.~C. Schuster, Experimental investigation of the al-y phase diagram, Journal of Alloys and Compounds 414~(1) (2006) 60--65.
\newblock \href {https://doi.org/10.1016/j.jallcom.2005.06.078} {\path{doi:10.1016/j.jallcom.2005.06.078}}.

\bibitem{devaraj2019}
A.~Devaraj, W.~Wang, R.~Vemuri, L.~Kovarik, X.~Jiang, M.~Bowden, J.~Trelewicz, S.~Mathaudhu, A.~Rohatgi, Grain boundary segregation and intermetallic precipitation in coarsening resistant nanocrystalline aluminum alloys, Acta Materialia 165 (2019) 698--708.

\bibitem{VASILIEV2004}
A.~Vasiliev, M.~Aindow, M.~Blackburn, T.~Watson, Phase stability and microstructure in devitrified al-rich al–y–ni alloys, Intermetallics 12~(4) (2004) 349--362.

\bibitem{RAGGIO2000}
R.~Raggio, G.~Borzone, R.~Ferro, The al-rich region in the y–ni–al system: microstructures and phase equilibria, Intermetallics 8~(3) (2000) 247--257.
\newblock \href {https://doi.org/10.1016/S0966-9795(99)00100-4} {\path{doi:10.1016/S0966-9795(99)00100-4}}.

\bibitem{Mika2010}
T.~Mika, B.~Kotur, Phase equlibria in the {Y, Gd}–ni–al ternary systems in the 65-100 at.\% al range at 773 k: a reinvestigation, Chemistry of Metals and Alloys 3 (2010).

\bibitem{zarechnyuk1980}
O.~Zarechnyuk, M.~Drits, R.~Rykhal, V.~Kinzhibalo, Examination of the mg-al-y system (0-33 at\% y) at 400◦ c, Metall 5 (1980) 242--244.

\bibitem{li2012}
X.~Li, M.~Starink, Dsc study on phase transitions and their correlation with properties of overaged al-zn-mg-cu alloys, Journal of Materials Engineering and Performance 21 (2012) 977--984.

\bibitem{LASA2002}
L.~Lasa, J.~Rodriguez-Ibabe, Characterization of the dissolution of the al2cu phase in two al–si–cu–mg casting alloys using calorimetry, Materials Characterization 48~(5) (2002) 371--378.
\newblock \href {https://doi.org/10.1016/S1044-5803(02)00283-8} {\path{doi:10.1016/S1044-5803(02)00283-8}}.

\bibitem{chen2020}
Z.~Chen, K.~Liu, E.~Elgallad, F.~Breton, X.-G. Chen, Differential scanning calorimetry fingerprints of various heat-treatment tempers of different aluminum alloys, Metals 10~(6) (2020) 763.

\bibitem{SHAO2023}
W.~Shao, J.~M. Guevara-Vela, A.~Fernández-Caballero, S.~Liu, J.~LLorca, Accurate prediction of the solid-state region of the ni-al phase diagram including configurational and vibrational entropy and magnetic effects, Acta Materialia 253 (2023) 118962.
\newblock \href {https://doi.org/10.1016/j.actamat.2023.118962} {\path{doi:10.1016/j.actamat.2023.118962}}.

\bibitem{janovszky2022}
D.~Janovszky, M.~Sveda, A.~Sycheva, F.~Kristaly, F.~Z{\'a}mborszky, T.~Koziel, P.~Bala, G.~Czel, G.~Kaptay, Amorphous alloys and differential scanning calorimetry (dsc), Journal of Thermal Analysis and Calorimetry (2022) 1--17.

\bibitem{laws2015}
K.~Laws, D.~Miracle, M.~Ferry, A predictive structural model for bulk metallic glasses, Nature communications 6~(1) (2015) 8123.

\bibitem{HUANG2019}
Z.~Huang, F.~Chen, Q.~Shen, L.~Zhang, T.~J. Rupert, Combined effects of nonmetallic impurities and planned metallic dopants on grain boundary energy and strength, Acta Materialia 166 (2019) 113--125.
\newblock \href {https://doi.org/10.1016/j.actamat.2018.12.031} {\path{doi:10.1016/j.actamat.2018.12.031}}.

\bibitem{miller2007}
M.~Miller, P.~Liaw, Bulk metallic glasses: an overview, Springer Science \& Business Media, 2007.

\bibitem{ZHANG2024}
X.~Zhang, Y.~Wan, C.~Chen, L.~Zhang, The effect of solute elements co-segregation on grain boundary energy and the mechanical properties of aluminum by first-principles calculation, Nanomaterials 14~(22) (2024).
\newblock \href {https://doi.org/10.3390/nano14221803} {\path{doi:10.3390/nano14221803}}.

\bibitem{OLMSTED2009}
D.~L. Olmsted, S.~M. Foiles, E.~A. Holm, Survey of computed grain boundary properties in face-centered cubic metals: I. grain boundary energy, Acta Materialia 57~(13) (2009) 3694--3703.
\newblock \href {https://doi.org/10.1016/j.actamat.2009.04.007} {\path{doi:10.1016/j.actamat.2009.04.007}}.

\bibitem{GARG2021}
P.~Garg, Z.~Pan, V.~Turlo, T.~J. Rupert, Segregation competition and complexion coexistence within a polycrystalline grain boundary network, Acta Materialia 218 (2021) 117213.
\newblock \href {https://doi.org/10.1016/j.actamat.2021.117213} {\path{doi:10.1016/j.actamat.2021.117213}}.

\bibitem{bober2016}
D.~B. Bober, A.~Khalajhedayati, M.~Kumar, T.~J. Rupert, Grain boundary character distributions in nanocrystalline metals produced by different processing routes, Metallurgical and Materials Transactions A 47 (2016) 1389--1403.

\bibitem{VANFLEET1995}
R.~R. Vanfleet, J.~Mochel, Thermodynamics of melting and freezing in small particles, Surface Science 341~(1) (1995) 40--50.
\newblock \href {https://doi.org/10.1016/0039-6028(95)00728-8} {\path{doi:10.1016/0039-6028(95)00728-8}}.

\bibitem{skripov1981}
V.~P. Skripov, V.~P. Koverda, V.~N. Skokov, Size effect on melting of small particles, physica status solidi (a) 66~(1) (1981) 109--118.
\newblock \href {https://doi.org/10.1002/pssa.2210660111} {\path{doi:10.1002/pssa.2210660111}}.

\bibitem{williams1996}
D.~B. Williams, C.~B. Carter, The transmission electron microscope, Springer, 1996.

\bibitem{tiryakioglu1998}
M.~Tiryakioglu, G.~Totten, Quenching aluminum components in water: Problems and alternatives, in: Heat Treating: Proceedings of the 18 th Conference, October 12 th-15 th, 1998, p.~3.

\bibitem{DUAN2024}
T.~Duan, W.~Kim, M.~Gao, J.~H. Perepezko, Crystallization of an undercooled zn-based glass forming alloy, Journal of Non-Crystalline Solids 627 (2024) 122823.
\newblock \href {https://doi.org/10.1016/j.jnoncrysol.2024.122823} {\path{doi:10.1016/j.jnoncrysol.2024.122823}}.

\bibitem{YANG2021}
Z.~Yang, R.~Al-Mukadam, M.~Stolpe, M.~Markl, J.~Deubener, C.~Körner, Isothermal crystallization kinetics of an industrial-grade zr-based bulk metallic glass, Journal of Non-Crystalline Solids 573 (2021) 121145.
\newblock \href {https://doi.org/10.1016/j.jnoncrysol.2021.121145} {\path{doi:10.1016/j.jnoncrysol.2021.121145}}.

\bibitem{Pogatscher2014}
S.~Pogatscher, P.~J. Uggowitzer, J.~F. Löffler, {In-situ probing of metallic glass formation and crystallization upon heating and cooling via fast differential scanning calorimetry}, Applied Physics Letters 104~(25) (2014) 251908.
\newblock \href {https://doi.org/10.1063/1.4884940} {\path{doi:10.1063/1.4884940}}.

\bibitem{neuber2020}
N.~Neuber, M.~Frey, O.~Gross, J.~Baller, I.~Gallino, R.~Busch, Ultrafast scanning calorimetry of newly developed au-ga bulk metallic glasses, Journal of Physics: Condensed Matter 32~(32) (2020) 324001.

\bibitem{CANTWELL2016}
P.~R. Cantwell, S.~Ma, S.~A. Bojarski, G.~S. Rohrer, M.~P. Harmer, Expanding time–temperature-transformation (ttt) diagrams to interfaces: A new approach for grain boundary engineering, Acta Materialia 106 (2016) 78--86.
\newblock \href {https://doi.org/10.1016/j.actamat.2016.01.010} {\path{doi:10.1016/j.actamat.2016.01.010}}.

\bibitem{davies1978}
H.~Davies, Rapid quenching techniques and formation of metallic glasses, Rapidly quenched metals III 1 (1978) 1--21.

\bibitem{LIAO2015}
J.~Liao, B.~Yang, Y.~Zhang, W.~Lu, X.~Gu, J.~Wang, Evaluation of glass formation and critical casting diameter in al-based metallic glasses, Materials \& Design 88 (2015) 222--226.
\newblock \href {https://doi.org/10.1016/j.matdes.2015.08.138} {\path{doi:10.1016/j.matdes.2015.08.138}}.

\bibitem{INOUE1998}
A.~Inoue, Amorphous, nanoquasicrystalline and nanocrystalline alloys in al-based systems, Progress in Materials Science 43~(5) (1998) 365--520.
\newblock \href {https://doi.org/10.1016/S0079-6425(98)00005-X} {\path{doi:10.1016/S0079-6425(98)00005-X}}.

\bibitem{GRAPES2017}
M.~D. Grapes, M.~K. Santala, G.~H. Campbell, D.~A. LaVan, T.~P. Weihs, A detailed study of the al3ni formation reaction using nanocalorimetry, Thermochimica Acta 658 (2017) 72--83.
\newblock \href {https://doi.org/10.1016/j.tca.2017.10.018} {\path{doi:10.1016/j.tca.2017.10.018}}.

\bibitem{JACKSON1999}
M.~Jackson, M.~Starink, R.~Reed, Determination of the precipitation kinetics of ni3al in the ni–al system using differential scanning calorimetry, Materials Science and Engineering: A 264~(1) (1999) 26--38.
\newblock \href {https://doi.org/10.1016/S0921-5093(98)01120-4} {\path{doi:10.1016/S0921-5093(98)01120-4}}.

\bibitem{yan2018}
F.~Yan, W.~Xiong, E.~Faierson, G.~B. Olson, Characterization of nano-scale oxides in austenitic stainless steel processed by powder bed fusion, Scripta Materialia 155 (2018) 104--108.

\bibitem{horn2022}
T.~Horn, C.~Rock, D.~Kaoumi, I.~Anderson, E.~White, T.~Prost, J.~Rieken, S.~Saptarshi, R.~Schoell, M.~DeJong, et~al., Laser powder bed fusion additive manufacturing of oxide dispersion strengthened steel using gas atomized reaction synthesis powder, Materials \& Design 216 (2022) 110574.

\bibitem{Khludkova1965}
A.~N. Khludkova, K.~V. Savitskii, Effect of quenching temperature on pore formation during the cyclic heat treatment of aluminum, Soviet Physics Journal 8 (1965) 19--22.

\bibitem{Bradshaw1957}
F.~J. Bradshaw, S.~Pearson, Quenching vacancies in aluminium, The Philosophical Magazine: A Journal of Theoretical Experimental and Applied Physics 2~(16) (1957) 570--571.
\newblock \href {https://doi.org/10.1080/14786435708243848} {\path{doi:10.1080/14786435708243848}}.

\bibitem{WERT1980}
J.~A. Wert, Vacancy concentrations in quenched binary alloys, Acta Metallurgica 28~(10) (1980) 1361--1373.
\newblock \href {https://doi.org/10.1016/0001-6160(80)90005-X} {\path{doi:10.1016/0001-6160(80)90005-X}}.

\bibitem{HUANG2023}
J.~Huang, M.~Li, J.~Chen, Y.~Cheng, Z.~Lai, J.~Hu, F.~Zhou, N.~Qu, Y.~Liu, J.~Zhu, Electronic structure and atomic migration of the fourth, fifth, and sixth period atoms in aluminum alloys: First principles calculation, Vacuum 210 (2023) 111823.
\newblock \href {https://doi.org/10.1016/j.vacuum.2023.111823} {\path{doi:10.1016/j.vacuum.2023.111823}}.

\bibitem{SWALIN1967}
R.~Swalin, C.~Yin, Thermal diffusion of vacancies in aluminum, Acta Metallurgica 15~(2) (1967) 245--248.
\newblock \href {https://doi.org/10.1016/0001-6160(67)90198-8} {\path{doi:10.1016/0001-6160(67)90198-8}}.

\bibitem{Banhart2020}
{Banhart, John}, {Yang, Zi}, {Liu, Meng}, {Madanat, Mazen}, {Zhang, Xingpu}, {Guo, Qianning}, {Yan, Yong}, {Röhsler, Andreas}, {Fricke, Konrad}, {Liang, Zeqin}, {Leyvraz, David}, {Hoell, Armin}, {Gericke, Eike}, {Wendt, Robert}, {Liu, Chunhui}, Exploring the hidden world of solute atoms, clusters and vacancies in aluminium alloys, MATEC Web Conf. 326 (2020) 01001.
\newblock \href {https://doi.org/10.1051/matecconf/202032601001} {\path{doi:10.1051/matecconf/202032601001}}.

\bibitem{milkereit2019}
B.~Milkereit, M.~J. Starink, P.~A. Rometsch, C.~Schick, O.~Kessler, Review of the quench sensitivity of aluminium alloys: analysis of the kinetics and nature of quench-induced precipitation, Materials 12~(24) (2019) 4083.

\bibitem{HARDY1954}
H.~Hardy, T.~Heal, Report on precipitation, Progress in Metal Physics 5 (1954) 143--278.
\newblock \href {https://doi.org/10.1016/0502-8205(54)90006-4} {\path{doi:10.1016/0502-8205(54)90006-4}}.

\bibitem{Lotter2018}
F.~Lotter, U.~Muehle, M.~Elsayed, A.~M. Ibrahim, T.~Schubert, R.~Krause-Rehberg, B.~Kieback, T.~E.~M. Staab, Precipitation behavior in high-purity aluminium alloys with trace elements – the role of quenched-in vacancies, physica status solidi (a) 215~(24) (2018) 1800375.
\newblock \href {https://doi.org/10.1002/pssa.201800375} {\path{doi:10.1002/pssa.201800375}}.

\bibitem{cahn1979}
J.~Cahn, Surface segregation in metals and alloys (1979).

\bibitem{weissmuller1993}
J.~Weissm{\"u}ller, Alloy effects in nanostructures, Nanostructured Materials 3~(1-6) (1993) 261--272.

\bibitem{chookajorn2012}
T.~Chookajorn, H.~A. Murdoch, C.~A. Schuh, Design of stable nanocrystalline alloys, Science 337~(6097) (2012) 951--954.
\newblock \href {https://doi.org/10.1126/science.1224737} {\path{doi:10.1126/science.1224737}}.

\bibitem{Cunningham2022}
W.~S. Cunningham, S.~T.~J. Mascarenhas, J.~S. Riano, W.~Wang, S.~Hwang, K.~Hattar, A.~M. Hodge, J.~R. Trelewicz, Unraveling thermodynamic and kinetic contributions to the stability of doped nanocrystalline alloys using nanometallic multilayers, Advanced Materials 34~(27) (2022) 2200354.
\newblock \href {https://doi.org/10.1002/adma.202200354} {\path{doi:10.1002/adma.202200354}}.

\bibitem{darling2013}
K.~Darling, A.~Roberts, Y.~Mishin, S.~Mathaudhu, L.~Kecskes, Grain size stabilization of nanocrystalline copper at high temperatures by alloying with tantalum, Journal of Alloys and Compounds 573 (2013) 142--150.
\newblock \href {https://doi.org/10.1016/j.jallcom.2013.03.177} {\path{doi:10.1016/j.jallcom.2013.03.177}}.

\bibitem{ZHOU2016}
N.~Zhou, T.~Hu, J.~Luo, Grain boundary complexions in multicomponent alloys: Challenges and opportunities, Current Opinion in Solid State and Materials Science 20~(5) (2016) 268--277.
\newblock \href {https://doi.org/10.1016/j.cossms.2016.05.001} {\path{doi:10.1016/j.cossms.2016.05.001}}.

\end{thebibliography}





\end{document}